\documentclass[pra,twocolumn,floatfix,notitlepage,groupedaddress,longbibliography]{revtex4-1}

\usepackage{mathtools}
\usepackage{scalerel,stackengine}
\stackMath
\newcommand\reallywidehat[1]{%
\savestack{\tmpbox}{\stretchto{%
  \scaleto{%
    \scalerel*[\widthof{\ensuremath{#1}}]{\kern-.6pt\bigwedge\kern-.6pt}%
    {\rule[-\textheight/2]{1ex}{\textheight}}
  }{\textheight}%
}{0.5ex}}%
\stackon[1pt]{#1}{\tmpbox}%
}
\usepackage{amssymb,amsmath,amstext,amsthm}
\usepackage{graphicx}
\usepackage{epstopdf}
\usepackage{color}
\usepackage[normalem]{ulem}
\usepackage{bm}
\usepackage{appendix}
\usepackage[T1]{fontenc}
\usepackage{bbm}
\usepackage{latexsym}
\usepackage{bbm}
\usepackage{float}
\usepackage{verbatim}
\usepackage{lipsum}
\usepackage{braket}
\usepackage{ulem}
\usepackage{mathtools}
\usepackage{mathrsfs}
 \usepackage{quantikz}
 \usepackage{dsfont}
\usepackage[colorlinks=false,citecolor=black]{hyperref}
\hypersetup{hidelinks}

\renewcommand{\leq}{\leqslant}

\newcommand{\ketbra}[2]{| \hspace{1pt} #1 \rangle \langle #2 \hspace{1pt} |}

\def \addPandAUNM {Department of Physics and Astronomy, University of New Mexico, Albuquerque, NM, USA}
\def \addCQuIC {Center for Quantum Information and Control, University
  of New Mexico, Albuquerque, NM, USA}
\def \addLANL {Theoretical Division, Los Alamos National Laboratory, Los Alamos, NM 87545, USA}

\begin{document}

\title{Spin-squeezed GKP codes for quantum error correction in atomic ensembles}

\author{Sivaprasad  Omanakuttan}
\email{somanakuttan@unm.edu}
\affiliation{\addCQuIC} \affiliation{\addPandAUNM}
\author{T.J.\,Volkoff}
\email{volkoff@lanl.gov}
\affiliation{\addLANL}

\begin{abstract}
GKP codes encode a qubit in displaced phase space combs of a continuous-variable (CV) quantum system and are useful for correcting a variety of high-weight photonic errors. 
Here we propose atomic ensemble analogs of the single-mode CV GKP code by using the quantum central limit theorem to pull back the phase space structure of a CV system to the compact phase space of a quantum spin system. 
We study the optimal recovery performance of these codes under error channels described by stochastic relaxation and isotropic ballistic dephasing processes using the diversity combining approach for calculating channel fidelity. 
We find that the spin GKP codes outperform other spin system codes such as cat codes or binomial codes. Our spin GKP codes based on the two-axis counter-twisting interaction and superpositions of SU(2) coherent states are direct spin analogs of the finite-energy CV GKP codes, whereas our codes based on one-axis twisting do not yet have well-studied CV analogs. 
A state preparation scheme for the spin GKP codes is proposed which uses the linear combination of unitaries method, applicable to both the CV and spin GKP settings. 
Finally, we discuss a fault-tolerant approximate gate set for quantum computing with spin GKP-encoded qubits, obtained by translating gates from the CV GKP setting using the quantum central limit theorem.
\end{abstract}
\maketitle

\section{Introduction\label{sec:intro}}
Quantum error correcting (QEC) codes in which logical quantum states are encoded by generating a comb structure in the phase space of a quantum oscillator \cite{PhysRevA.64.012310} are central to proposals for fault-tolerant quantum computation with continuous variables (CV) \cite{Bourassa2021,Larsen2021,PhysRevA.101.012316,PhysRevX.8.021054,Raveendran2022finiterateqldpcgkp}. 
In the ideal (infinite energy) limit, the code states of a single mode Gottesman-Kitev-Preskill (GKP) qubit are defined by a superposition of quadrature eigenvectors, forming a Dirac comb with peaks displaced by $\sqrt{2\pi}$ along a phase space direction. These CV GKP codes protect perfectly against sufficiently small shifts in momentum and position (within the half-period $\sqrt{\pi/2}$) although, 
importantly, high-performance error correction properties of the CV GKP code extend beyond Gaussian coherent errors. For example, the CV GKP code outperforms a variety of bosonic codes for the case of photon loss \cite{PhysRevA.97.032346}. 
There have been multiple proposals for implementations and applications of the CV GKP codes in bosonic systems \cite{Bourassa2021,xu2022qubit,noh2020encoding}, and low-energy code states have been experimentally demonstrated  \cite{de2022error}. 
Codes similar to GKP motivated by the specific nature of the errors in the physical system could be fundamental in building fault-tolerant quantum computation with lower overhead, e.g., with fewer number of physical systems comprising the logical qubits \cite{puri2020bias,wu2022erasure, PhysRevX.12.021049,ofek2016extending}.

Similar to CV systems, atomic spin systems are potential candidates for large scale quantum information processing due to their controllability and potential for entanglement generation, both crucial properties for  quantum computational tasks which require a large number of qubits.
Quantum simulations with more than $200$ qubits have been experimentally demonstrated with neutral atoms \cite{ebadi2021quantum,scholl2021quantum}. 
Such large atomic ensembles have found interesting applications in the context of quantum metrology and large-scale entanglement generation. Proposals for QEC codes native to spin systems utilizing conversion of the relevant error to erasure errors and exploiting spin system symmetries have also been proposed recently \cite{wu2022erasure,Gross2021}.
Therefore, the potential for utilizing a collective spin degree of freedom to robustly encode a logical qubit has been recognized. However, in this work, we fully utilize the analogies between the overcomplete basis of SU(2) coherent states and CV coherent states, and between atomic spin squeezing interactions and CV nonlinear interactions to explore how the CV GKP qubit can be ported to a spin-$N/2$ system. The analogy is a two-way street, and we further expect that some of the spin-$N/2$ codes introduced in this work have useful CV analogues based on higher-order effective photon-photon interactions.

Our families of spin GKP codes are built around the concept of spin squeezing \cite{PhysRevA.47.5138}, which is the compact phase space analogue of quadrature squeezing used to define finite-energy GKP states for CV systems. 
The technical result that allows us to port the CV GKP codes to the spin-$N/2$ system is the quantum central limit theorem, which we phrase in terms of asymptotic isomorphism of Lie algebras under an appropriate Hilbert space isometry.
Using this technique, the spin-$N/2$ two-axis countertwisting interaction allows defining spin GKP codes depending on the two-body interaction strength in a similar way as finite-energy CV GKP codes depend on the squeezing parameter \cite{PhysRevA.47.5138}. We also introduce spin GKP codes based on the one-axis twisting interaction and based on superpositions of coherent states on a phase space grid.
For both noise models, we consider, the spin GKP codes are optimized for the best recovery performance using an efficient semi-definite programming (SDP) approach known as the  diversity combining method  \cite{PhysRevA.75.012338}.

One of the significant challenges in using GKP states for quantum computation, regardless of whether one considers the spin-$N/2$ or CV setting, is that the preparation of the code states is not straightforward. 
Multiple approaches to GKP state generation have been proposed using the quantum phase estimation algorithm and using Kerr interaction.
In this work, we propose to use the recently developed linear combination of unitaries method \cite{berry2015simulating,low2019hamiltonian,holmes2022quantum}, which is now a prominent module for quantum algorithms.
This scheme is  applicable, in principle, to general grid states, although the size of the grid is limited by the controllability of an ancilla degree of freedom. To implement universal, fault-tolerant quantum computation with spin-$N/2$ GKP-encoded qubits, we require rotation-resistant gates, and we develop these gates using the quantum central limit theorem to port the displacement resistant gates from the CV GKP setting to the spin-$N/2$ GKP setting.
The fault-tolerant approximate Clifford group generators are composed of Hamiltonians at most quadratic in angular momentum operators of two spin-$N/2$ systems. Assuming that these generators can be implemented with high fidelity, the magic states sufficient for universal quantum computation
can be generated using a similar approach as in  \cite{PhysRevLett.123.200502}.

The remainder of this article is organized as follows. In Section \ref{sec:motivation} we discuss how the quantum central limit theorem motivates spin analogues of the CV GKP codes.  
In Section \ref{sec:spingkp} we introduce a family of spin  GKP code states using both two-axis countertwisting and one-axis twisting atomic interactions, and we compare the free parameters that appear in these codes to parameters that define finite-energy CV GKP codes.
In Section \ref{sec:errors_performance} we consider two kinds of error channels relevant to atomic ensembles and compare the recovery performance of our various versions of spin GKP codes to that of other spin-$N/2$ bosonic codes. The main result is that the spin GKP codes perform better than other codes over a large range of noise strengths for both of the channels we consider. 
In Section \ref{sec:gkp_prep_LCU} we present an algorithm to prepare the GKP states based on the linear combination of unitaries technique  and discuss potential schemes for creating spin GKP states using the light-matter interaction.
Lastly, in Section  \ref{sec:universal_gate_set} we describe  how one could construct a universal gate set for the spin GKP codes following again the quantum central limit theorem.
Discussions and possible future directions are in Section \ref{sec:discussions}.

The appendices are devoted to important technical details.
In App.\ref{sec:app_diversity_combining}, we describe the diversity combining approach which allows us to efficiently calculate the optimal recovery channel.
 App.\ref{app:sbp} provides details about the stochastic relaxation channel and App.\ref{sec:Numerical_Benchmarking_details} discusses numerical benchmarking details for optimal recovery with respect to the stochastic relaxation and isotropic ballistic dephasing channels. In App.\ref{sec:Ballistic_dephasing_along_one_axis}, the one-axis ballistic dephasing channel is discussed in order to show that other collective spin QEC codes can outperform spin GKP codes when the noise is known to be directional.
Lastly, in App.\ref{sec:app}, we analyze the optimal recovery of various finite-energy CV GKP codes under photodetection noise, in order to show that different well-motivated versions of the GKP code can have different recovery performances.

\section{Motivation from quantum central limit theorem and spin squeezing\label{sec:motivation}}
In this work, we provide several schemes to encode a logical qubit in a collection of $N$ qubits which differ fundamentally from the current experimental approaches for fault-tolerant quantum computation which require individual addressing of each qubit. Specifically, our logical states can be viewed as symmetric atomic ensembles in which the $N$ qubits are prepared in the symmetric subspace and addressed by global operations. 
This system is mathematically equivalent to a spin-$j$ system where $j=N/2$.
In this section, we describe the motivation of the encoding schemes in detail.

The logical state encodings we detail in Section \ref{sec:spingkp} are motivated by combining spin squeezing of a symmetric $N$-qubit ensemble with the GKP scheme for encoding logical qubits in CV systems.
Finite-energy (i.e., physical) versions of the ideal CV GKP code are constructed by forming approximate eigenvectors of the stabilizer operators which are CV displacement operators.
These stabilizers define the phase space lattice structure of the code projector.  
In a spin-$N/2$ representation of SU(2), as occurs in, e.g., a system of $N$ indistinguishable bosonic atoms distributed among two orthogonal modes, a strictly canonical approach based on stabilizers would lead to the replacement of idealized CV quadrature eigenstates in the original CV GKP code by idealized position eigenstates $\ket{(\theta,\phi)}$ on the Riemann sphere \cite{PhysRevX.10.031050}.
However, if one wants to work with physical states of the symmetric $N$-qubit system and make use of realizable interactions such as spin squeezing, it is unclear how to construct logical states defining a code projector associated with a lattice-like phase space distribution allowing for isotropic correction of small rotation errors. 
  The approach taken in this work generates the characteristic comb structure of CV GKP logical states in a system of $N$ bosons by taking superpositions of rotated spin-squeezed states and utilizing the asymptotics of these states with respect to atom number $N$ to make a connection with the phase space structure of CV GKP codes. This connection has its roots in the Cushen-Hudson quantum central limit theorem.

In modern terminology, part of the original Cushen-Hudson quantum central limit theorem \cite{ch} states that the characteristic function $\text{tr}\rho^{\otimes M}e^{ix\bar{q}_{M}+iy\bar{p}_{M}}$ of the multimode canonical operators $\bar{q}_{M}:={1\over\sqrt{M}}\sum_{j=1}^{M}q_{j}$, $\bar{p}_{M}:={1\over\sqrt{M}}\sum_{j=1}^{M}p_{j}$ in the $M$-mode CV state $\rho^{\otimes M}$ converges pointwise on $\mathbb{R}^{2}$ to the Gaussian characteristic function $\text{tr}e^{ixq+ipy}\rho_{G}$ of the single mode CV state $\rho_{G}$, the Gaussification of $\rho$ \cite{PhysRevLett.123.050501}. This convergence was characterized in Ref.\cite{Becker2021}. Physically, this theorem says that quantum fluctuations of quadratures of $M$ CV modes can be described by a Gaussian quantum state. Subsequently, inspired by mathematical treatments of laser theory, Giri and von Waldenfels extended this part of the Cushen-Hudson central limit theorem to the free Lie algebra generated by an associative algebra \cite{Giri1978}. For the purposes of the present work, it is sufficient to have convergence of (spin-squeezed) SU(2) coherent states in the spin-$N/2$ representation to (quadrature squeezed) Heisenberg-Weyl coherent states of a single CV modes. Accardi and Bach utilized the argument of \cite{Giri1978} applied to spin generators to prove precisely this convergence in Theorems 4.2 and 6.2 of \cite{ab}.

To explicitly state how we use the limits of \cite{ab}, we first relate the  orthonormal basis of Dicke states for the symmetric subspace of $N$ qubits to the  Fock basis of a single mode CV system using an isometry $V$. Specifically, we define $V:\ket{N-k,k}\rightarrow \ket{k}$, where the Dicke state $\ket{N-k,k} := {1\over \sqrt{{N\choose k}}}\sum_{\text{Ham}(x)=k}\ket{x_{1}}\cdots \ket{x_{N}}$, $x_{j}\in \lbrace 0,1\rbrace$. For large $N$, we have the following relations between the Schwinger boson spin operators $J_{+}:=a^{\dagger}_{0}a_{1}$, $J_{-}=J_{+}^{\dagger}$, $J_{z}={1\over 2}(a^{\dagger}_{0}a_{0} - a_{1}^{\dagger}a_{1})$ under this isometry:
\begin{align}
    Ve^{{\alpha\over \sqrt{N}}J_{-} -{\overline{\alpha}\over \sqrt{N}}J_{+}}\ket{N,0}  \rightarrow e^{\alpha a^{\dagger}-\overline{\alpha}a}\ket{0}\nonumber \\
    Ve^{{z\over N} J_{+}^{2} - {\overline{z}\over N}J_{-}^{2}}\ket{N,0}  \rightarrow e^{z a^{2}-\overline{z}a^{\dagger 2}}\ket{0}
    \nonumber \\
    Ve^{i\theta J_{z}}\ket{N-k,k} = e^{-i\theta a^{\dagger}a}\ket{k}
    \label{eq:isometric_equations}
\end{align}
where the limits are in the weak topology on the Hilbert space of the harmonic oscillator and the last equality is only valid in the projective Hilbert space.
These relations imply that a $1/\sqrt{N}$-scaled spin rotation generator about an axis in the $xy$-plane is asymptotically equivalent to the generator of CV displacement $\alpha a^{\dagger}-\overline{\alpha}a$, and that the $1/N$-scaled generator of two-axis countertwisting (TACT) is asymptotically equivalent to the generator of quadrature squeezing.
Note that straightforwardly applying a Holstein-Primakoff transformation to the left hand sides of the first two lines of (\ref{eq:isometric_equations}) results in $N$-dependent algebraic functions of the creation and annihilation operators in the exponents on the right hand side of (\ref{eq:isometric_equations}). Such parameterized generators do not naturally appear in finite-energy approximations to the ideal CV GKP code. In Section \ref{sec:spingkp}, our first spin-$N/2$ GKP code $\texttt{tactgkp}$ is defined by pulling back (using the isometry $V$) a finite-energy CV GKP code defined by superposing displaced quadrature squeezed states with Gaussian weights. This is the form of the finite-energy CV GKP code originally proposed  in \cite{PhysRevA.64.012310}.

The equivalence of defining ideal CV GKP codes from a superposition of displaced quadrature eigenvectors or from a lattice superposition of CV coherent states motivates a spin-$N/2$ code that consists of an approximately Gaussian weighted superposition of SU(2) coherent states.
Our spin-$N/2$ codes $\texttt{spinGKP}$ and $\texttt{unigkp}$ are defined in \eqref{eqn:spingkpdef} and \eqref{eqn:unigkp}, respectively, and both codes asymptote to a square lattice finite-energy CV GKP code in the $N\rightarrow \infty$ limit. The optimal recovery properties of these codes must be analyzed in addition to those of $\texttt{tactgkp}$ because the equivalence of their respective CV asymptotics does not imply that the spin-$N/2$ counterparts exhibit equivalent optimal recovery properties under spin system noise channels. We consider realistic noise channels that apply to atomic ensembles independent of $N$, viz., the noise channels we consider are actual decoherence processes for atomic ensembles, not simply pullbacks of the commonly considered CV bosonic attenuation channel by $V$. Further, the comparison is especially relevant due to the fact that  an ensemble of $O(100)$ atoms represents the present state-of-the-art system size for spin squeezing.
For CV GKP codes, optimization of the defining lattice is known to give order-of-magnitude higher channel fidelity for small noise parameters \cite{PhysRevA.97.032346}. In the present work, we focus on schemes for generating logical states of a spin-$N/2$ atomic ensemble possessing comb-like distributions of $J_{x}$ or $J_{y}$, while leaving the question of optimization of the phase space ``lattice'' of the corresponding code projectors as a direction for future investigation (``lattice'' in quotations because the phase space is spherical). We expect that even for finite $N$, such optimization of the codes by utilizing results on packing of spherical caps or spherical ellipses will further increase the error recovery performance. 

Lastly, we will also introduce in Section \ref{sec:spingkp} a class of spin-$N/2$ GKP codes that are not  asymptotically equal to finite-energy CV GKP codes as $N\rightarrow \infty$. The construction of these codes is enabled by the existence of spin squeezing generators that do not involve $J_{\pm}$.  For example, instead of utilizing the TACT generator, one can seek to generate the characteristic GKP comb structure of the code projector by utilizing the one-axis twisting (OAT) interaction $J_{z}^{2}$. Our OAT-based spin-$N/2$ codes are $\texttt{oatGKP}$ and $\texttt{oatGKP(uni)}$, which differ from each other by the form of the amplitudes of the rotated OAT parent state. The OAT interaction is mathematically analogous to the Kerr interaction in a single-mode CV system, with both interactions generating multipronged cat states for specific constant interaction times. We emphasize that all of the codes in Section \ref{sec:spingkp} define finite energy single-mode CV states in the $N\rightarrow \infty$ limit.

For the codes $\lbrace \texttt{tactgkp}, \texttt{spingkp}, \texttt{oatgkp}\rbrace$, we define amplitudes on the rotated spin-squeezed states that asymptotically approach the Gaussian amplitudes in finite-energy CV GKP codes. While the exact form of such amplitudes can be informed by specific aspects of the experimental state generation protocol, for simplicity we define the amplitudes by utilizing the deMoivre-Laplace formula in the following form, which holds for $x\in \mathbb{R}$:
\begin{align}
    {\Gamma(N+1)\over 2^{N}\Gamma({N\over 2}+x+1)\Gamma({N\over 2}-x+1)} &\sim \sqrt{2\over \pi N}e^{-2x^{2}/N}.
    \label{eq:dml}
\end{align}

\section{Spin-$N/2$ GKP codes\label{sec:spingkp}}

In this section, we provide definitions for the spin-$N/2$ GKP codes whose quantum error correction properties we explore in this work. The kinematic structure of the code states is motivated by considering a comb-like structure in phase space, similar to that in CV GKP code states. The CV GKP codes are designed to allow correction of small displacements of the $q$ and $p$ quadratures, and the spin-$N/2$ GKP states are analogously designed to allow correction of small $\mathrm{SU}(2)$ rotations generated by $J_{y}$ and $J_{x}$. We find in Section \ref{sec:errors_performance} that the phase space structure of the spin-$N/2$ code states allows high-fidelity optimal recovery with respect to specific noise channels and also noise-agnostic recovery.

\begin{figure}
    \centering
    \includegraphics[width=\columnwidth]{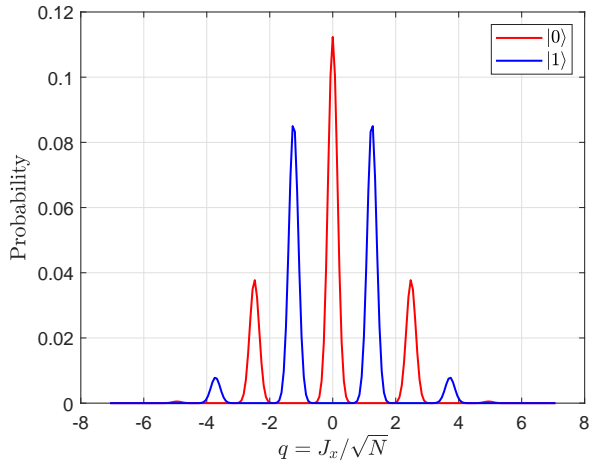}
    \caption{\textbf{Probability distribution of the codewords in the $q$ basis.} 
    The probability distribution of the  \texttt{tactgkp} for $N=100$ , $T=10$, and $\delta=0.3$ is shown in the basis of eigenvectors of $J_x/\sqrt{N}$, which asymptotes to the CV quadrature $q$ under isometry (\ref{eq:isometric_equations}). 
    Similar to the CV GKP codes one can see the well-separated combs which are central to the error correction properties of the grid state codes.}
    \label{fig:figure_probability_distribution_x}
\end{figure}

The logical states of our spin-$N/2$ qubits will be written as $\ket{\mu_{S}}$, where $\mu$ is the qubit orthonormal basis index and $S$ is a list of real parameters. Later in the paper, when we fix the parameters of a certain spin-$N/2$ GKP code for various purposes, we will omit $S$ and simply label the logical states as $\ket{\mu_{L}}$ with $L$ standing for ``logical''. The first spin-$N/2$ GKP code strictly converges to a finite-energy CV GKP code by combining the quantum central limit theorem and the deMoivre-Laplace theorem. This code is labeled $\texttt{tactgkp}$ and has the exact expression
\begin{widetext}
\begin{equation}
\ket{\mu_{\lambda,z}}\propto \left[\sum_{t=1}^{T}\Gamma(N+1) \left(\frac{e^{-2i(\sqrt{2\pi}t+\mu \sqrt{\pi\over 2}) {J_{y}\over \sqrt{N}} }}{  C^{N,t}_{\lambda,\mu}  C^{N,t}_{-\lambda,\mu}}+\frac{e^{2i(\sqrt{2\pi}t-\mu \sqrt{\pi\over 2}) {J_{y}\over \sqrt{N}} }}{  C^{N,t}_{\lambda,-\mu}C^{N,t}_{-\lambda,-\mu} }\right)+\Gamma(N+1)\frac{e^{-2i\mu \sqrt{\pi\over 2} {J_{y}\over \sqrt{N}} }}{C^{N,0}_{\lambda,\mu}C^{N,0}_{-\lambda,\mu}}\right] e^{{z\over 2N}(J_{+}^{2} -J_{-}^{2})}\ket{N,0}
   \label{eqn:spingkp1} 
\end{equation}
\end{widetext}
where to lighten the notation we use,
\begin{equation}
    C^{N,t}_{\lambda,\mu}=\Gamma\left({N\over 2}+\lambda\sqrt{\pi N}\left(t+{\mu\over 2}\right)+1\right),
\end{equation}
 with $\mu \in \lbrace 0,1\rbrace$ and $\ket{N,0}=\ket{J,J_{z}=J}$ is the spin coherent state with maximal $J_{z}$ eigenvalue. By considering the spectral projections of the operator $J_{x}/\sqrt{N}$, which is asymptotically equivalent to the CV operator $q$ under the isometry $V$ in (\ref{eq:isometric_equations}), one can obtain a probability distribution analogous to  the $q$-quadrature distribution of a CV GKP code. In Fig.~\ref{fig:figure_probability_distribution_x}, we show this distribution for the case of $N=100$, $T=10$, and $\delta=0.3$.
 
 In (\ref{eqn:spingkp1}), there are three $N$-indexed sequences that contribute to the large $N$ convergence of the code states to CV GKP code states: 1. the sequence of $J_{y}$ rotations converges to a displacement of CV quadrature $q$, 2. the sequence of TACT states converges to a $q$-quadrature squeezed state, and 3. the sequence of coefficients of each $J_{y}$ rotation converges to a Gaussian weight. The convergence of the first and second sequences follows from the quantum central limit theorem, according to which the $2/\sqrt{N}$ prefactor of the rotation generators causes convergence of the rotations about $J_{y}$ to CV displacements of $q$ with the same parameter, while the $1/N$ prefactor of the TACT interaction causes convergence of spin squeezing with interaction time $z=\tanh^{-1}{d-1\over d+1}$ to CV squeezing of the canonical $q$ quadrature with squeezing parameter $z$ and covariance matrix ${1\over 2}\text{diag}(d^{-1},d)$. 
 For consistency with common convention in the CV literature, we use $\lambda= \delta$ and $d= \delta^{-2}$, which allows the code states to be parametrized with a single parameter, but one can also choose the value of $d$ independently from $\lambda$ based on minimization of the overlap between the code states or based on best optimal recovery for a specific noise channel. 
Lastly, the convergence of the coefficient of each $J_{y}$ rotation to a Gaussian weight follows from \eqref{eq:dml}. Specifically, the three coefficients asymptote to
$e^{-2\pi\lambda^{2}(t\pm {\mu \over 2})^{2}}$ and $e^{-{\pi\lambda^{2}\over 2}\mu^{2}}$, respectively, up to a $\mu$-independent prefactor. The asymptotically Gaussian amplitudes suggest that the fidelity between code states should be insensitive to contributions from $t\ge 1/2\pi \lambda^{2}$ in the sum index, and  we empirically find that taking $T= 5$ is appropriate for the range of $\lambda$ that gives a low fidelity between codewords.  
 
In a spin-$N/2$ system, finite-energy considerations do not demand that a superposition of discretely rotated spin-squeezed states have decaying coefficients with respect to changing the rotation index from zero. Therefore, it is valid to consider a simplified version of \texttt{tactgkp} defined solely as a superposition of rotated two-axis countertwisted SU(2) coherent states  \cite{PhysRevA.47.5138}. We refer to this code as $\texttt{tactgkp(uni)}$ for its uniform amplitudes on the rotated spin-squeezed states (which become more distinguishable as $N$ is increased), and the code states are explicitly given by
 \begin{equation}
     \ket{\mu_{z}}\propto \sum_{t=-T}^{T}e^{2i(\sqrt{2\pi}t-\mu \sqrt{\pi\over 2}) {J_{y}\over \sqrt{N}} } e^{{z\over 2N}(J_{+}^{2} -J_{-}^{2})}\ket{N,0} 
     \label{eq:tact_gkp}
 \end{equation}
 with $\mu \in \lbrace 0,1\rbrace$ giving the two code space and 
 \begin{equation}
 z=\tanh^{-1}{\delta^{-2}-1\over \delta^{-2}+1}
 \label{eq:sq}
 \end{equation}
 is the squeezing parameter. 
 Unlike the three-parameter $\texttt{tactgkp}$, the $\texttt{tactgkp(uni)}$ code states are a function of only the TACT interaction time $z$ and the superposition cut off $T$ and one could optimize both of these two in order to get the best performance.
 The $T$ parameter can be determined by fixing the particle number and noise model and optimizing the channel fidelity over $z$ for each $T$.

For single-mode CV GKP codes, the connection between a Gaussian-weighted superposition of quadrature squeezed states and a Gaussian-weighted superposition of coherent states on a lattice in phase space $\mathbb{R}^{2}$ can be made due to the Gaussian nature of quadrature squeezed states \cite{PhysRevA.97.032346}. Similar to the CV GKP code defined by a superposition of CV coherent states with mean vectors on a square grid in phase space \eqref{eqn:sqgridcv}, we define a spin-$N/2$ code called \texttt{spingkp}
\begin{widetext}
    \begin{equation}
    \vert \mu_{\delta}\rangle \propto  \sum_{t\in \lbrace -T,\ldots, T\rbrace^{\times 2}}\frac{\Gamma(N+1)}{ D^{N,t_1,t_2}_{\delta,\mu} D^{N,t_1,t_2}_{-\delta,\mu}}e^{-2i(\sqrt{2\pi}t_{1}+\mu{\sqrt{\pi\over 2}}){J_{y}\over \sqrt{N}}}e^{2i\sqrt{\pi\over 2}t_{2}{J_{x}\over \sqrt{N}}}\vert N,0\rangle
    \label{eqn:spingkpdef}
\end{equation}
\end{widetext}
where to lighten the notation we use,
\begin{equation}
    D^{N,t_1,t_2}_{\delta,\mu}=\Gamma\left({N\over 2} + {\delta \over 2}\sqrt{N\pi ((2t_{1}+\mu)^{2}+t_{2}^{2})} +1 \right),
\end{equation}
with $\mu \in \lbrace 0,1\rbrace$. If one removes the amplitudes, one gets the code \texttt{unigkp} which depends only on the superposition cutoff $T$, and is given by
\begin{equation}
    \vert \mu \rangle \propto \sum_{t\in \lbrace -T,\ldots, T\rbrace^{\times 2}}
e^{-2i(\sqrt{2\pi}t_{1}+\mu{\sqrt{\pi\over 2}}){J_{y}\over \sqrt{N}}}e^{2i\sqrt{\pi\over 2}t_{2}{J_{x}\over \sqrt{N}}}\vert N,0\rangle.
\label{eqn:unigkp}
\end{equation}

Unlike quadrature squeezing in a single-mode CV system, spin squeezing in a spin-$N/2$ system can be realized in many ways depending on the two-body interaction that is implemented. For example, one-axis twisting \cite{PhysRevA.47.5138} is one of the fundamental interactions exploited in atomic systems for quantum simulation and quantum metrology. This interaction can be used to create GKP code states

\begin{widetext}
\begin{equation}
\ket{\mu_{\lambda,z}}\propto \left[\sum_{t=1}^{T}\Gamma(N+1) 
\left(\frac{e^{-2i(\sqrt{2\pi}t+\mu \sqrt{\pi\over 2}) {J_{y}\over \sqrt{N}} }}{  C^{N,t}_{\lambda,2\mu}  C^{N,t}_{-\lambda,2\mu}}
+\frac{e^{-2i(\sqrt{2\pi}t-\mu \sqrt{\pi\over 2}) {J_{y}\over \sqrt{N}} }}{  C^{N,t}_{\lambda,-2\mu}C^{N,t}_{-\lambda,-2\mu} }\right)
+\Gamma(N+1)\frac{e^{-2i\mu \sqrt{\pi\over 2} {J_{y}\over \sqrt{N}} }}{C^{N,0}_{\lambda,2\mu}C^{N,0}_{-\lambda,2\mu}}\right]e^{-i\delta J_{x}^{2}}\ket{N,0}
   \label{eqn:oatgkp2} 
\end{equation}
\end{widetext}
where  $\mu \in \lbrace -{1\over 2},{1\over 2}\rbrace$.  Although the combinatorial prefactors go to Gaussians as $N\rightarrow \infty$, for the present code there is no \textit{a priori} relation between the parameters $\delta$ and $\lambda$ that can be imposed as was done for $\texttt{tactgkp}$ with the aim of convergence to a finite energy CV GKP code. Therefore, we analyze two reduced parameter versions of (\ref{eqn:oatgkp2}): \texttt{oatGKP} having unparameterized, nonuniform amplitudes 
\begin{widetext}
\begin{equation}
    \ket{\mu_{\delta}}\propto \left[\sum_{t=1}^{T} {N\choose {N\over 2}+t}\left[ e^{-2i(\sqrt{2\pi}t+\mu \sqrt{\pi\over 2}) {J_{y}\over \sqrt{N}} } + e^{2i(\sqrt{2\pi}t-\mu \sqrt{\pi\over 2}) {J_{y}\over \sqrt{N}} } \right]+ {N\choose N/2}e^{-2i\mu \sqrt{\pi\over 2} {J_{y}\over \sqrt{N}} }\right]e^{-i\delta J_{x}^{2}}\ket{J,J_\mathrm{z}=J}
\label{eqn:oatgkp}
\end{equation}
\end{widetext}

and \texttt{oatGKP(uni)} with uniform amplitudes on rotated one-axis twisted states
\begin{align}
    \ket{\mu_{\delta}}\propto& \;  \sum_{t=-T}^{T}e^{2i\left(\sqrt{2\pi}t-\mu\sqrt{\frac{\pi}{2}}\right)\frac{J_\mathrm{y}}{\sqrt{N}}} e^{-i\delta J_{x}^{2}}\ket{J,J_{\mathrm{z}}=J}
\label{eqn:oatgkp3}
\end{align}
where $\mu \in \lbrace 1/2,-1/2\rbrace$ for both codes.   We use the \texttt{oatgkp} and \texttt{oatgkp(uni)}     defined in Eq.~(\ref{eqn:oatgkp}) and Eq.~(\ref{eqn:oatgkp3}) in this article.   Note that the $\delta \rightarrow 0$ limit ($\delta \rightarrow 1$ limit) corresponds to infinite interaction time (zero interaction time) for \texttt{tactgkp} (\texttt{oatgkp}).                   

Taking the logical index as $\mu \in \lbrace 1/2,-1/2\rbrace$ instead of $\mu\in\lbrace 0,1\rbrace$ effectively shifts the code states in phase space and, for finite $N$, has implications for optimal recovery performance with respect to the noise channels considered in this work. In fact, a similar point arises when requiring CV GKP codes states to have exactly the same number of photons in expectation (see Appendix \ref{sec:app} for a short review of square grid CV GKP codes, equal-energy versions of these codes, and their recovery properties). For the equal-energy CV GKP codes, we observed slightly reduced optimal recovery performance in the CV case.


\begin{figure*}
    \centering
    \includegraphics[width=\textwidth]{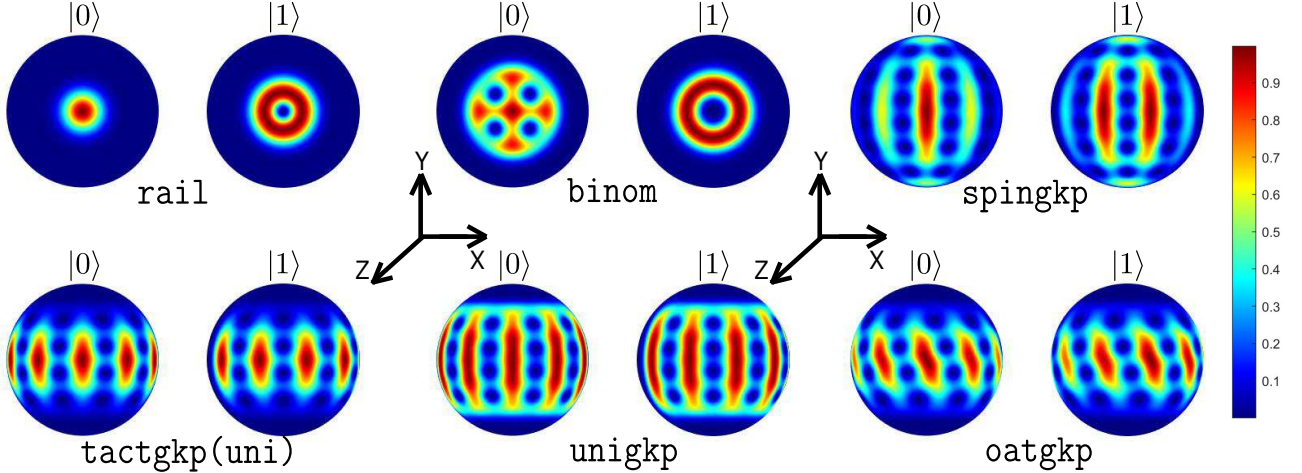}
    \caption{\textbf{Husimi Q function.} The  spin Husimi Q function, $Q(\alpha)=\frac{1}{\pi}\langle\alpha|\hat{\rho}|\alpha\rangle$ where $\ket{\alpha}$ is the spin-$N/2$ coherent state, for the code states considered in this article.
    For the spin version of the GKP codes we have derived in this article, we can see the comb-like structure originating in the phase space similar to the comb-like structure in phase space for CV GKP codes. 
    The parameters used for the simulation of \texttt{spingkp} is $\delta=0.2$ and $T=5$, for \texttt{tactgkp(uni)} is $\delta=0.3$ and $T=2$, for \texttt{unigkp} is $T=2$, and for \texttt{oatgkp} is $\delta=0.04$ and $T=5$. } 
    \label{fig:husimi_code_words}
\end{figure*}

The Husimi Q representation is one of the well-known  quasiprobability distributions used to represent the phase space structure of a quantum state of a CV system. 
A spin analog of the Husimi Q representation is defined by the analogy of CV coherent states to SU(2) coherent states \cite{agarwal2012quantum}.
In Fig.~\ref{fig:husimi_code_words}, we show the  spin-$N/2$ Husimi Q function $Q(\alpha)=\frac{1}{\pi}\langle\alpha|\hat{\rho}|\alpha\rangle$ where $\ket{\alpha}$ is a spin coherent state, for the main code states considered in this article. 
For the spin GKP codes, a distorted comb-like structure is seen to appear in spherical phase space, with TACT interaction producing longitudinal alignment of the rotated spin-squeezed states and OAT interaction producing rotated spin-squeezed states aligned on twisted axes.

For comparison to the spin GKP codes, we consider classes of well-known bosonic codes. The binomial code (\texttt{binom}) \cite{PhysRevX.6.031006} is defined by
\begin{equation}
\begin{aligned}
 \ket{0_{L}}&=\frac{\ket{J=\frac{N}{2},J_z=\frac{N}{2}}+\ket{J=\frac{N}{2},J_z=\frac{N}{2}-4}}{\sqrt{2}}\\
 &=\frac{\ket{N,0}+\ket{N-4,4}}{\sqrt{2}},\\
 \ket{1_{L}}&=\ket{J=\frac{N}{2},J_z=\frac{N}{2}-2}\\
 &=\ket{N-2,2},
\end{aligned}
   \end{equation}
and the single-rail encoding \cite{RevModPhys.79.135} \texttt{rail} code which is defined by $\ket{0_{L}}=\ket{N,0}$ and $\ket{1_{L}}=\ket{N-1, 1}$.

One can also define the spin-$N/2$ cat qubit codes $\texttt{cat}$ as,
\begin{equation}
\begin{aligned}
\ket{0}_{\vec{n}}&={\ket{J,\vec{n}\cdot \vec{J}=J}+\ket{J,\vec{n}\cdot \vec{J}=-J}\over \sqrt{2}}\\
\ket{1}_{\vec{n}}&= { \ket{J,\vec{n}\cdot \vec{J}=J}-\ket{J,\vec{n}\cdot \vec{J}=-J} \over \sqrt{2}}
    \label{eq:cat_state}
\end{aligned}
\end{equation}
where $\vec{n}$ is the specific orientation of interest defined by the $(\theta,\phi)$ in the spherical coordinates. 
Also, we have the coherent state codes \texttt{coherentstate} given as,
\begin{equation}
\begin{aligned}
\ket{0}_{\vec{n}}&=\ket{J,\vec{n}\cdot \vec{J}=J}\\
\ket{1}_{\vec{n}}&=\ket{J,\vec{n}\cdot \vec{J}=-J}
    \label{eq:coherent_state}
\end{aligned}
\end{equation}
where, similar to $\texttt{cat}$,  $\vec{n}$ is the specific orientation of interest defined by the $(\theta,\phi)$ in the spherical coordinates.
One can similarly port other CV coherent-state-based codes from the bosonic system to the spin system.

\begin{figure}
    \centering
    \includegraphics[width=0.48\textwidth]{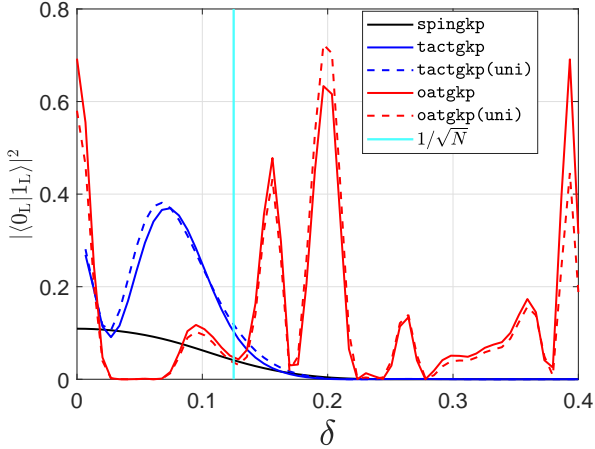}
    \caption{\textbf{$\delta$ parameter and code state quasi-orthogonality.} Fidelity is given as a function of $\delta$ between the two codewords for \texttt{tactgkp}, \texttt{spingkp}, \texttt{oatgkp}, and \texttt{oatgkp(uni)} with $N=64$.  We observe a wide range of potential choice of $\delta$ giving quasi-orthogonal code states, which can be used as an additional degree to control according to the noise channel of interest. 
    Compared to other codes, the codes \texttt{tactgkp} and\texttt{spingkp} exhibit a larger domain of $\delta$ corresponding to a well-defined comb-like structure, which is due to the different physical origin of these codewords. The \texttt{tactgkp} code uses TACT to generate the comb-like structure, whereas \texttt{oatgkp} and \texttt{oatgkp(uni)} use OAT. The line $y=1/\sqrt{N}$ is shown to indicate the $\delta$ value above which $\texttt{oatgkp}$ no longer has the comb-like structure necessary for high-fidelity recovery.}
    \label{fig:fidelity_figure}
\end{figure}

Unlike the CV GKP codes, the spin squeezing parameter $z$ of the spin-$N/2$ GKP codes is merely kinematic and not related to the number of particles comprising the code states. It depends on the strength of the interatomic interaction used to generate the code states. For CV GKP codes, the desired code energy constrains the value of the squeezing parameter  \cite{PhysRevA.97.032346}.
However, fixing the relation between $z$ and $\delta$ as above for spin-$N/2$ GKP codes, we can identify a range of valid $\delta$ based on minimization of the overlap between code states. 
In Fig.~\ref{fig:fidelity_figure}, the overlap between the two code states $\ket{0_{L}},\ket{1_{L}}$ is shown as a function of $\delta$ for the \texttt{spingkp} and \texttt{oatgkp} is shown. For the case  of \texttt{spingkp}, there is a large domain of $\delta$ where quasi-orthogonality holds, whereas for the case of \texttt{oatgkp} the domain of quasi-orthogonality is smaller and disconnected.
Note that if a range of energies is allowed for finite energy versions of the CV GKP states, consideration of quasi-orthogonality could also enter into a choice for the $\delta$ parameter.

Finally, note that in Ref.\cite{PhysRevA.97.032346}, optimization over a subset of lattices in single-mode CV phase space $\mathbb{R}^{2}$ was performed in order to identify the best geometric configuration for optimal recovery properties.
Similar analyses can be carried out for the spin-$N/2$ codes introduced in this section, but we now  focus on examining optimal recovery properties under two noise channels relevant for large spin systems.

\section{Optimal code performance under ensemble errors\label{sec:errors_performance}}

We now define two noise channels and compare the optimal recovery performance of the spin-$N/2$ GKP codes discussed in the previous section. 
The optimal recovery properties of the codes are  analyzed according to the following standard procedure: 1. two logical states $\ket{0_{L}}$ and $\ket{1_{L}}$ are defined according to the $\mu$ parameter for the various encodings discussed in the previous section, 2. a noise channel $\mathcal{N}_{\xi}$ of interest is applied for a subset of valid noise strengths $\xi$, and 3. the optimal recovery performance is quantified according to the optimal channel fidelity 
\begin{align}
F_{\mathcal{N}_{\gamma}}=\max_{\mathcal{R}}\langle \psi_{\rho}\vert (\mathcal{R}\circ \mathcal{N}_{\xi}) \otimes \mathbf{1} \left( \ket{\psi_{\rho}}\bra{\psi_{\rho}} \right) \vert \psi_{\rho} \rangle 
\label{eqn:rtrt}
\end{align}
where $\ket{\psi_{\rho}}$ is a purification of the state $\rho={1\over 2}P_{\text{code}}$, $P_{\text{code}}=\ket{0_{L}}\bra{0_{L}} +\ket{1_{L}}\bra{1_{L}}$. With the code states being in a Hilbert space of dimension $N+1$, one can show that (\ref{eqn:rtrt}) can be solved using an SDP with a positive $(N+1)^{2}\times (N+1)^{2}$ matrix variable \cite{PhysRevA.75.012338}. Even using efficient SDP solvers like SeDuMi \cite{sturm1999using}, this optimization is computationally expensive  for $N=O(10)$.
An alternative and more computationally efficient approach is obtained by considering the encoding and noise channel as a single noise channel from a qubit system to the spin-$N/2$ system and the recovery channel as a recovery from the spin-$N/2$ system to the qubit system. This ``diversity combining'' approach \cite{PhysRevA.75.012338} allows us to analyze optimal recovery for spin-$N/2$ GKP codes with $N=O(10)$ quickly on a laptop computer and is discussed in more detail in Appendix \ref{sec:app_diversity_combining}.

In the first two subsections of this section, we apply the diversity combining method to a scenario in which one has perfect encoding, recovery, and decoding and the only error we assume is the one arising from the specific noise channel considered. 
Thus the fidelity (\ref{eqn:rtrt}) that we calculate provides a theoretical bound on the performance of these codes for a specific noise channel. Note that the corresponding optimal recovery channels obtained do not necessarily correspond to an experimentally feasible design. Lastly, in the third subsection, we introduce a  noise-agnostic recovery that is intended to work for any weak noise channel or compositions thereof.

The noise channels considered in Section \ref{sec:src}, \ref{sec:bdc} may not prove to be the most relevant ones for the experimental implementations, however, to our knowledge, there is not a widely agreed upon set of typical noise models for atom ensemble-based control, unlike the situation in, e.g., quantum optics.
Thus, instead of considering more general noise models, we find it instructive to focus on two noise channel corresponding to different physical processes (atom loss and dephasing, respectively) and understand the optimal recovery performance of the spin-$N/$ GKP codes under these decohering processes.


\subsection{Stochastic relaxation channel\label{sec:src}}
 The single-mode quantum photon detection process
\begin{equation}
    \rho \mapsto \mathcal{N}_{\gamma}(\rho):= \sum_{\ell=0}^{\infty}{(1-e^{-\gamma})^{\ell}\over \ell!}e^{-{\gamma\over 2} a^{\dagger}a}a^{\ell}\rho a^{\dagger \ell}e^{-{\gamma\over 2} a^{\dagger}a}
    \label{eq:photon_detection_process}
\end{equation}
is a quantum channel describing random photon loss in both free-space and waveguide-based photonics \cite{Ueda_1989}. We consider a spin-$N/2$ analog of the above process, which describes the stochastic relaxation of a two-level atom system to a uniform ground state. To get the atomic version, define orthonormal internal states $\ket{0}$ and $\ket{1}$ and assume that $\ket{0}$ has lower energy. Analogous to the no-count photonic process defined in Ref.\cite{Ueda_1989}, the no-count atomic detection process is $\mathcal{S}^{(0)}_{\tau}(\rho_{t})=e^{-{\lambda \tau\over 2}(N-2J_{z})}\rho_{t} e^{-{\lambda \tau\over 2}(N-2J_{z})}$ (note that $N-2J_{z}=a_{1}^{\dagger}a_{1}$, so the no-count process results in an exponential decay of the amplitudes for $\ket{N-m,m}$ with $m>0$). This process is analogous to the backaction photon decay in (\ref{eq:photon_detection_process}). Analogous to the one-count, instantaneous photon detection process, we define an instantaneous relaxation that transfers an atom from $\ket{1}$ to $\ket{0}$, and one can describe its action on the rank-1 operators $\ket{N-m,m}\bra{N-m',m'}$ as 
\begin{align}
    &{}\mathcal{S}^{(1)}(\ket{N-m,m}\bra{N-m',m'})\nonumber \\
    &{} ={\lambda J_{+}\ket{N-m,m}\bra{N-m',m'}  J_{-}\over \sqrt{(N-m+1)(N-m'+1)}} 
    \label{eq:jump_operators}.
\end{align} 
The prefactor can be interpreted as an occupation-dependent scaling of the probability of an atom decaying from $\ket{1}$ to $\ket{0}$.  Using the fact that $e^{{\lambda \over 2}J_{z}}J_{+}^{k}e^{-{\lambda \over 2}J_{z}} = e^{{k\lambda\over 2}}J_{+}^{k}$, we obtain the full stochastic relaxation channel
\begin{equation}
    T_{t}(\rho)=\sum_{\ell=0}^{N}{(1-e^{-\lambda t})^{\ell}\over \ell!\lambda^{\ell}}e^{-{\lambda t\over 4} (N-2J_{z})}(\mathcal{S}^{(1)})^{\ell}(\rho)e^{-{\lambda t\over 4} (N-2J_{z})}
    \label{eqn:oppp}
\end{equation}
which one can verify to be completely positive and trace-preserving. The trace-preserving property is shown explicitly in Appendix \ref{app:sbp}.

The stochastic relaxation channel is a generalization of the amplitude damping channel for the qubit, similar to how the CV detection loss channel is a generalization of single-photon attenuation dynamics \cite{7442587}.
This can be seen from the Eq.~\eqref{eq:jump_operators}, where the action of the decoherence process is the action of the ladder operator $J_{+}$, which increases the $J_z$ projection, lowering the energy of the ensemble. 
The steady state of the decoherence process is the uniform state in which each spin is in the ground state, thus generalizing the single-qubit amplitude damping channel. 

\begin{figure*}
    \centering
    \includegraphics[width=0.9\textwidth]{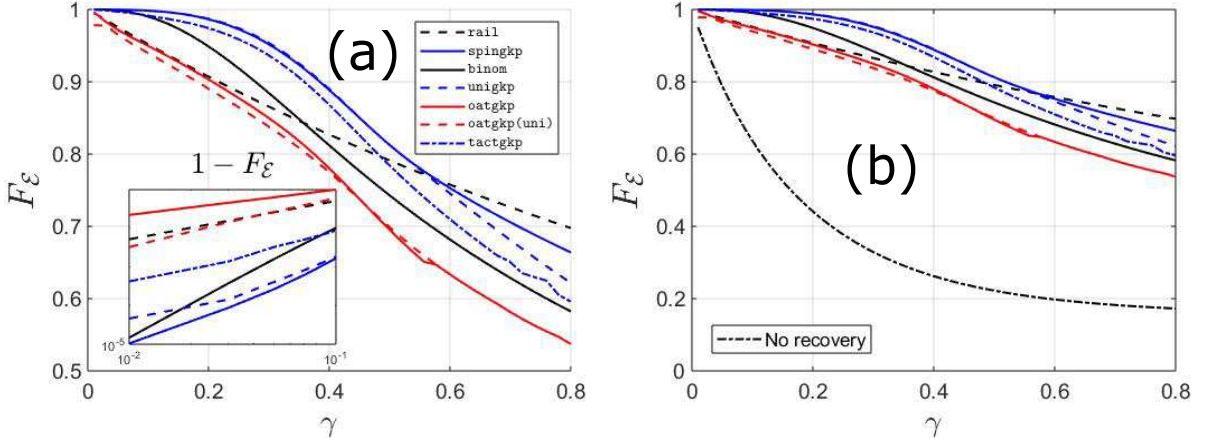}
    \caption{\textbf{Channel fidelity for stochastic relaxation.} In (a) the channel fidelity is given as a function of $\gamma$ for the noise model $\mathcal{E}=\mathcal{N}_{\gamma}$ given in Eq.~\eqref{eqn:oppp} for $N=64$ and cut off $T=5$ for all codes except \texttt{unigkp} and \texttt{oatgkp (uni)} where we all optimize the cut off value too. 
    The inset in the figure gives the channel infidelity as a function of $\gamma$ for the small values of $\gamma$ suggesting that the GKP codes outperform the binomial code and rail code for a large range of $\gamma$ and thus showing its potential advantage compared to other codes for spin systems for stochastic relaxation channel.
    In (b) the average fidelity for the \texttt{spingkp} without any recovery for $20$ random states are given for reference to show the fact the recovery is crucial.}
    \label{fig:stochastic_relaxation}
\end{figure*}

The performance of the various code-words under the stochastic noise channel is given in Fig.~\ref{fig:stochastic_relaxation}, one could see that the  GKP codes outperform the binomial code and rail code for a large range of $\gamma$. 
These results are very similar to the results for the CV GKP codes in \cite{PhysRevA.102.032408} for the case of the photon loss case.
In the inset, we show the small $\gamma$ behavior to further illustrate the behavior of these codes. 
This numerical data indicate the usefulness of the GKP states based on TACT or displaced coherent states for stochastic relaxation, similar to known results for the CV GKP states with respect to amplitude damping.
\begin{figure*}
    \centering
    \includegraphics[width=\textwidth]{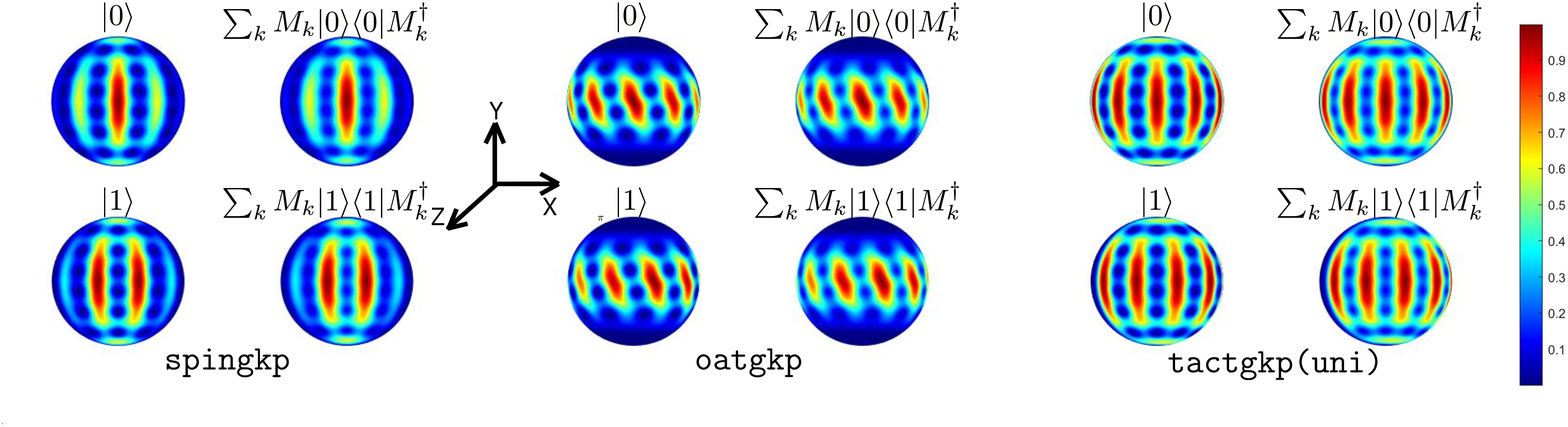}
    \caption{\textbf{Code states under stochastic relaxation.} Husimi distribution of the code states for the \texttt{oatgkp} and \texttt{spingkp} for the quantum channel considered in Eq.~(\ref{eqn:oppp}) for $\gamma=0.2$. 
    The plots are shown for $\delta=0.2,T=5$,  $\delta=0.04, T=5$, and $\delta=0.2,T=2$ respectively for the \texttt{spingkp},\texttt{oatgkp}, and \texttt{tactgkp(uni)}.
    The states $\sum_{k}M_k\ketbra{0}{0}M_k^{\dagger}$ and $\sum_{k}M_k\ketbra{1}{1}M_k^{\dagger}$ are the states one obtains after the action of the decoherence channel for the state $\ket{0}$ and $\ket{1}$ respectively.}
    \label{fig:stochastic_relaxation_husimi}
\end{figure*}
However, the \texttt{oatgkp} performs worse than all other channels considered here and we account for this fact that the error channel destroys the comb-like structure and creates superposition between the combs which is the crucial ingredient in recovering errors for the GKP kind of states. 
This also underlies the significance of comb structure for the GKP states and the effect of these codewords for  $\gamma=0.2$ is given in Fig.~\ref{fig:stochastic_relaxation_husimi}. 
The effect of the channel for these two codewords are significantly different and the distinguishability of the combs in the \texttt{oatgkp} is lost whereas the \texttt{spingkp} still preserves the comb-like structure even after the action of the quantum channel.
This manifests in the better recovery of the \texttt{spingkp} compared to the \texttt{oatgkp} as given in Fig.~\ref{fig:stochastic_relaxation}.
The results can also be interpreted by examining the underlying physical interactions creating these states, which  limits the range of $\delta$ that define quasiorthogonal code states for the \texttt{oatgkp} (see Fig.~\ref{fig:fidelity_figure}).

\subsection{Ballistic dephasing channel\label{sec:bdc}}
Ballistic dephasing along a single spin axis is known to destroy the spin-squeezing generated by one-axis twisting of an orthogonal spin direction \cite{baamara2021squeezing}.
Such a one-axis ballistic dephasing channel can be written \cite{PhysRevLett.115.010404}
\begin{equation}
\Phi_{\sigma}(\rho) \propto \int_{-\pi}^{\pi}d\theta \, e^{-{1\over 2\sigma}\theta^{2}} e^{-i\theta J_{z}} \rho e^{i\theta J_{z}}.
\label{eqn:ballistic_dephasing}
\end{equation}
where the $x$-axis is chosen just for illustration.
More details about this channel are given in App.~(\ref{sec:Ballistic_dephasing_along_one_axis}).

Here we consider  an isotropic ballistic dephasing channel, a generalized version of the one-axis ballistic dephasing channel,  defined as
\begin{equation}
\Phi_{\sigma}(\rho)\propto \int_{-\pi}^{\pi}d\theta \int_{0}^{\pi/2}d\xi \int_{0}^{2\pi}d\phi \, \sin\xi\,  e^{-{\theta^{2}\over 2\sigma}}e^{-i\theta \vec{n}\cdot \vec{J}}\rho e^{i\theta \vec{n}\cdot \vec{J}}
\label{eq:isotropic_ballistic}
\end{equation}
where $\vec{n}=(\sin \xi \cos \phi,\sin \xi \sin \phi,\cos \xi)$ is a unit vector.
We focus on the isotropic case because the isotropic phase space structure of the spin-$N/2$ GKP codes is an advantage in this case compared to codes that are defined relative to a specific axis, such as \texttt{cat}. 
One can view this channel, which delocalizes whatever information isotropically in all directions as can be seen from Fig.~(\ref{fig:ballistic_isotropic_husimi}) 

\begin{figure*}
    \centering
    \includegraphics[width=\textwidth]{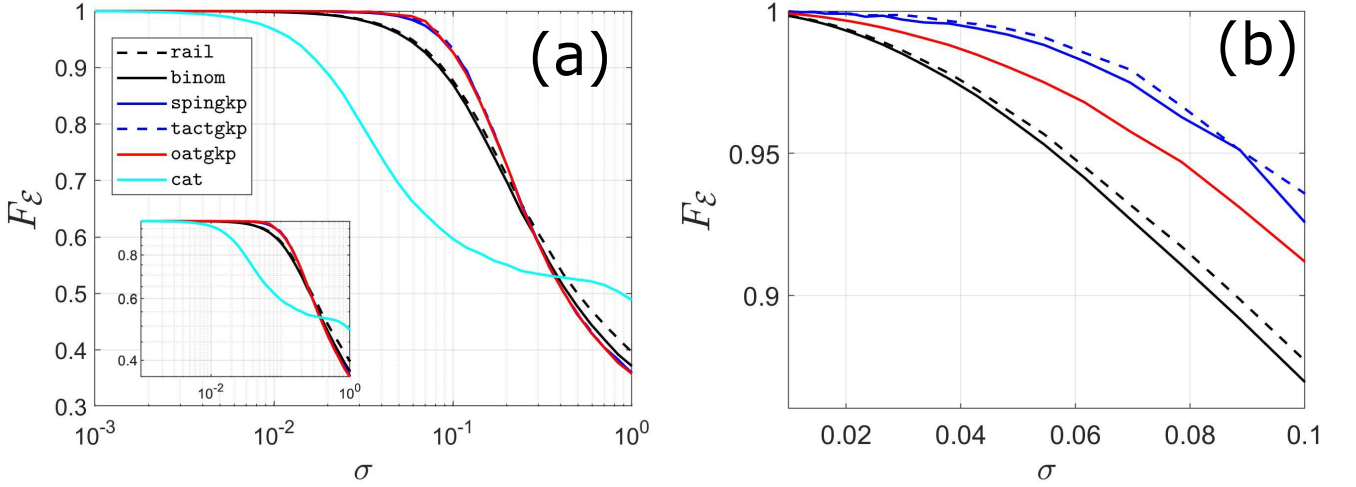}
    \caption{\textbf{Channel fidelity for isotropic ballistic dephasing.} (a) The channel fidelity as a function of $\sigma$ for the noise model given in Eq.~\eqref{eq:isotropic_ballistic} for $N=64$.
    The inset figure is the same data given in a log scale for $\sigma$.
    From the figure, one can infer that the \texttt{spingkp} and \texttt{oatgkp} outperforms all other codes.
    In (b) the case of small values of $\sigma$ is given to further illustrate the performance of different spin-$N/2$ GKP code states. Thus in the very low error limit \texttt{tactgkp} outperforms other codes for small noise, and \texttt{oatgkp} outperforms other codes for intermediate noise values}
    \label{fig:ballistic_isotropic}
\end{figure*}

The noise strength regime $\sigma \gtrsim N^{-2}$ is where the optimal recovery of different spin-$N/2$ codes can be usefully compared. 
The optimal recovery results for the isotropic ballistic dephasing are shown in Fig.~\ref{fig:ballistic_isotropic} for $N=64$ (see Appendix \ref{sec:Numerical_Benchmarking_details} for code parameters associated with the optimized data in the figure).
From the figure, one can see that the \texttt{spingkp} and \texttt{oatgkp} states outperform all other codes for the ballistic dephasing channel.
When the dephasing is only along one axis, we find that \texttt{cat} outperforms all other codes we considered (see Appendix \ref{sec:Ballistic_dephasing_along_one_axis} for detailed discussion). However, when the noise is isotropic, \texttt{cat} is no longer the optimal code.
\begin{figure*}
    \centering
    \includegraphics[width=\textwidth]{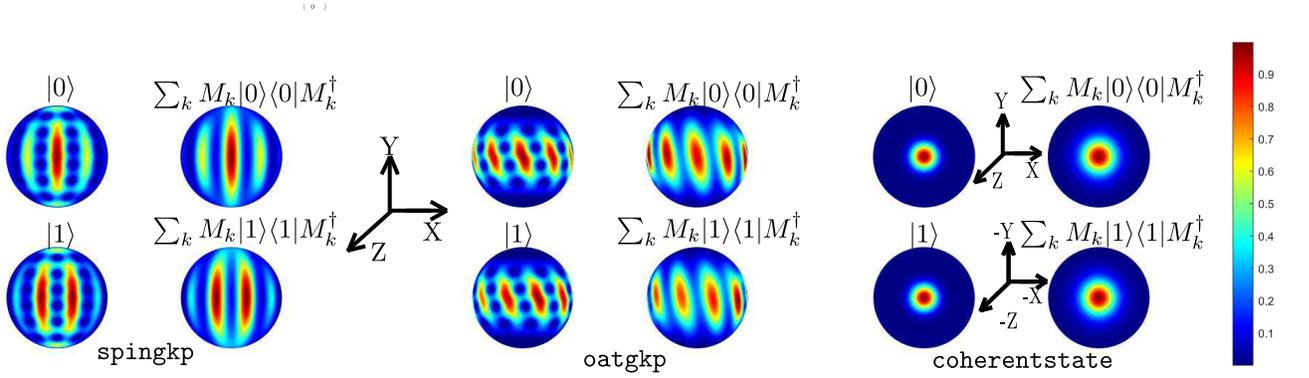}
    \caption{\textbf{Code states under isotropic ballistic dephasing.} Husimi distribution of the code states for the \texttt{oatgkp} and \texttt{spingkp} for the quantum channel considered in Eq.~(\ref{eqn:oppp}) for $\sigma=0.2$. 
    The plots are shown for $\delta=0.2,T=5$,  $\delta=0.04, T=5$, respectively for the \text{spingkp}  and \texttt{oatgkp}.
    The effect of the channel is to isotropically displace the state as can be seen clearly from the case of the coherent state. 
    For the case of the spin-$N/$ GKP codes $\texttt{oatgkp}$ and $\texttt{spingkp}$, the comb-like structure is preserved which helps in the recovery of the ideal code states.
    This in turn leads to the better recovery fidelity (see Fig.~\ref{fig:ballistic_isotropic}) for the GKP based codes compared to the other codes.}
    \label{fig:ballistic_isotropic_husimi}
\end{figure*}

To further understand the nature of the isotropic ballistic dephasing channel, the Husimi distribution of the code states for the \texttt{oatgkp} and \texttt{spingkp} are shown in Fig.~\ref{fig:ballistic_isotropic_husimi} for noise strength $\sigma=0.2$.
    The effect of the channel is to isotropically widen the phase space support of the code states, as can be seen clearly from the case of the coherent state and from the general equation
    \begin{align}
\text{Var}_{\Phi_{\sigma}(\rho)}\vec{m}_{1}\cdot \vec{J}&=(1-{2\sigma \over 3})\text{Var}_{\rho}\vec{m}_{1}\cdot \vec{J} \nonumber \\
&{} +{\sigma\over 3}\langle (\vec{m}_{2}\cdot \vec{J})^{2}\rangle_{\rho}+{\sigma\over 3}\langle (\vec{m}_{3}\cdot \vec{J})^{2}\rangle_{\rho}+O(\sigma^{2})
\end{align}
where $\vec{m}_{1,2,3}$ is an orthonormal basis of $\mathbb{R}^{3}$.
    This behavior is reminiscent of the thermal noise bosonic Gaussian channel in the CV setting. 
    For the $\texttt{oatgkp}$ and $\texttt{spingkp}$  code states, the comb-like structure is preserved which allows high optimal channel fidelity even for moderate noise strengths.

Although a specific recovery channel could be used to obtain a lower bound for the channel fidelity for the noise channel (\ref{eq:isotropic_ballistic}), one can show that, for mixed unitary channels, the correctable part of the QEC kernel is a rigorous lower bound for the channel fidelity. Consider a quantum channel of the form $\rho \mapsto \mathcal{E}(\rho):= \mathbb{E}_{U\sim p}U\rho U^{\dagger}$ where the expectation is taken with respect to some probability density $p(U)$ on the unitary group, and define the QEC  kernel by
\begin{align}
    \epsilon(U,U')&= a_{U,U'}P_{\text{code}} +b_{U,U'}X_{\text{code}} \nonumber \\
    &{} + c_{U,U'}Y_{\text{code}} + d_{U,U'}Z_{\text{code}}
    \label{eqn:qeckern}
\end{align}
where the coefficient kernel $a(U,U')={1\over 2}\text{tr}P_{\text{code}}U^{\dagger}U'$ is designated as the correctable part. In other words, the QEC kernel takes as input two unitary operators and outputs a 2$\times$2 matrix. Note that channel (\ref{eq:isotropic_ballistic}) has the mixed unitary form. We show that the expected square of the correctable part of the QEC matrix provides a lower bound for the optimal channel fidelity. One simply considers the recovery channel 
\begin{equation}
    \mathcal{R}(\cdot)=R_{1}(\cdot)R_{1}^{\dagger} + R_{2}(\cdot) R_{2}^{\dagger}
\end{equation}
with $R_{1}=S^{\dagger}U^{\dagger} $ and $R_{2}$ is any other Kraus operator that completes the channel and notes that
\begin{align}
    \text{max}_{\mathcal{R}} \sum_{j}\mathbb{E}_{V\sim p}\vert \text{tr}R_{j}VS\rho' \vert^{2} 
    &\ge \text{max}_{U}\mathbb{E}_{V\sim p}\vert \text{tr}R_{1}VS\rho' \vert^{2}\nonumber \\
    &= \text{max}_{U}\mathbb{E}_{V\sim p}\vert \text{tr}S^{\dagger}U^{\dagger}VS\rho' \vert^{2}\nonumber \\
    &\ge {1\over 4}\mathbb{E}_{(U,V)\sim p^{\times 2}}\big\vert \text{tr}U^{\dagger}VP_{\text{code}} \big\vert^{2} \nonumber \\
    &\ge  {1\over 4}\big\vert\mathbb{E}_{(U,V)\sim p^{\times 2}} \text{tr}U^{\dagger}VP_{\text{code}} \big\vert^{2}
    \label{eqn:rgrg}
\end{align}
\begin{figure}
    \centering
    \includegraphics[width=0.48\textwidth]{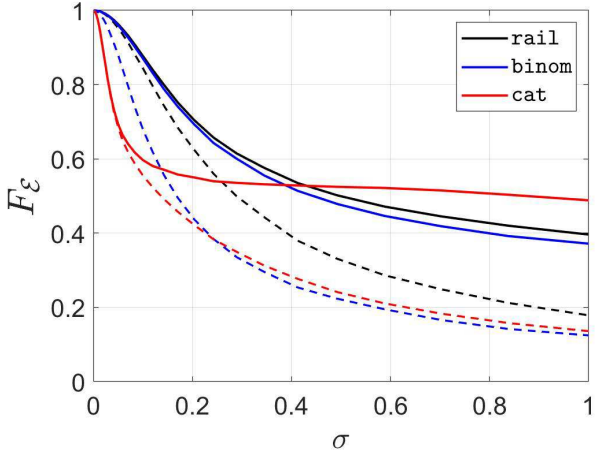}
    \caption{\textbf{Channel fidelity lower bounds for isotropic ballistic dephasing.} 
    Comparison of the channel fidelity obtained by the optimal recovery map for the ballistic isotropic channel (solid lines) to the lower bounds on the recovery map obtained in the second line of the Eq.(\ref{eqn:rgrg}) (color-corresponding dashed lines).
}
    \label{fig:ballistic_istropic_1}
\end{figure}
where we refer to (\ref{eqn:rbrb1}) for the notation $\rho'$ and $S$. We also note that the first two inequalities in (\ref{eqn:rgrg}) still hold when $\rho'$ is replaced by an arbitrary linear operator on $\mathbb{C}^{2}$.  Since for small noise values, the trivial recovery channel is optimal, we expect the right-hand side of (\ref{eqn:rgrg}) to give an indication of $F_{\mathcal{E}}$ at small noise strengths. Similarly, a rapidly vanishing expected non-correctable part with respect to decreasing noise indicates a recovery rate rapidly approaching 1 for small noise. The benefit of analyzing (\ref{eqn:rgrg}) as a lower bound to the optimal channel fidelity is that it does not require one to identify a cleverly chosen recovery map. In Fig.~\ref{fig:ballistic_istropic_1}, we compare the lower bound in the second line of the (\ref{eqn:rgrg}) to the maximal channel fidelity obtainable by recovery map for the ballistic isotropic channel.

\subsection{Noise Agnostic Recovery for Spin GKP}
\label{sec:nar}
We have seen that the GKP codes for spin systems outperform other codes for optimal recovery of specific error channels. However, neither the optimal recovery channel solving (\ref{eqn:rtrt}) nor the optimal unitary recovery in (\ref{eqn:rgrg}) are necessarily implementable in practice since they are obtained by optimizing over large sets of channels acting on the symmetric subspace.
Here we introduce a recovery process that is defined irrespective of any specific noise model (i.e., noise-agnostic), and is motivated by the comb-like structure of the GKP codes. 
Similar to the continuous variable GKP recovery process, the spin analog also follows from two steps as given in the following circuits,
\begin{figure}[!ht]
    \centering
    \includegraphics[width=0.42\textwidth]{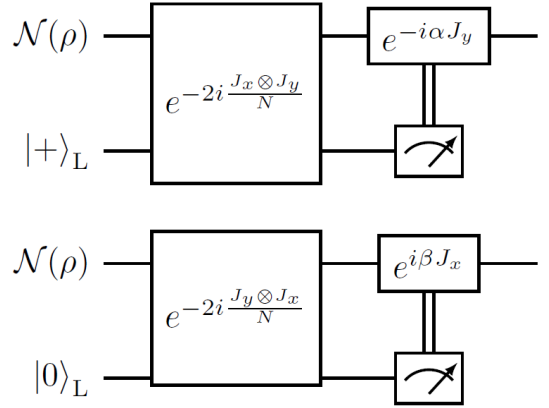}
    \caption{circuit for noise agnostic recovery}
    \label{fig:recovery_circuit}
\end{figure}
To understand how this recovery process works, we can focus on the rotation error in the $y$ direction. 
First, note that the $\ket{+}_{\mathrm{L}}\propto \ket{0}_{\mathrm{L}}+\ket{1}_{\mathrm{L}}$ has an approximate comb structure with angular spacing $\sqrt{2\pi \over N}$ on the circle  defined at latitude $y$.
An interaction between the spin ensembles given by $e^{-2i{J_x\otimes J_{y}\over N}}$ can be used to approximately imprint the rotation error onto an ancillary spin ensemble prepared in $\ket{+}_{\mathrm{L}}$.
Thus one could recover the state as long as the error is small. Similarly, for large $N$, the state $\ket{0}_{\mathrm{L}}$ has an approximate comb structure with angular spacing $\sqrt{2\pi \over N}$ on the circle defined at latitude $x$ and again using the interaction $e^{-2i{J_x\otimes J_{y}\over N}}$ can be used to approximately imprint the rotation error onto an ancillary spin ensemble.

In the CV setting, one carries out a homodyne measurement to diagnose the error. Here we introduce an analog of the homodyne measurement for the spin-$N/2$ system defined by the  positive operator-valued measurement containing the rank one projections onto 
\begin{equation}
\lbrace \ket{\psi_{z}(a)}:=e^{-iaJ_{y}}e^{{z\over 2N}(J_{+}^{2}-J_{-}^{2})}\ket{N,0}\rbrace_{a=-\pi}^{\pi}
\label{eqn:povm}
\end{equation}
(if one is considering correcting errors on \texttt{tactgkp}) allows to obtain a syndrome $a$, from which a residue $R_{\sqrt{2\pi/N}}(a)\in [-\sqrt{\pi/2N},\sqrt{\pi/2N}]$ is computed from $R_{\sqrt{2\pi/N}}(a)=a-\lambda_{a}$, where $\lambda_{a}$ is the element of $\lbrace - \lfloor \sqrt{N\pi/2}\rfloor,\ldots,-\sqrt{2\pi /N},0,\sqrt{2\pi/N},\ldots,\lfloor \sqrt{N\pi/2}\rfloor\rbrace$ nearest to $a$. Note that the measurement (\ref{eqn:povm}) is the spin-$N/2$ analog of the position measurement for CV GKP correction of $q$-displacement errors, but that for finite $N$, one must complete the measurement with the positive operator $\mathbb{I}-\Pi_{z}$ where $\Pi_{z}$ is the projection onto the span of the $\ket{\psi_{z}(a)}$ for all $a$.

\begin{figure}
    \centering
    \includegraphics[width=\columnwidth]{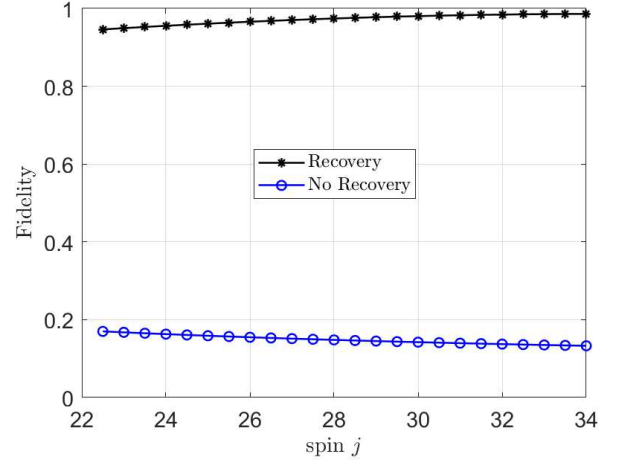}
    \caption{\textbf{Performance of the noise agnostic recovery.} 
    Comparison of the fidelity of the recovered state with the state is given in  Eq.~\eqref{eq:initial_state} for a rotation error $e^{ibJ_y}$ with $b=\sqrt{2\pi/N}/4$.
    The recovered fidelity is calculated by identifying the syndrome with the maximum probability for the first part of the recovery circuit given in Fig.~(\ref{fig:recovery_circuit}).
    For larger values of $j$, the recovery operation results in larger fidelity, indicative of the success of our noise-agnostic recovery circuit (\ref{fig:recovery_circuit}).
}
    \label{fig:noise_agnostic_recovery_1}
\end{figure}

To see the performance of noise agnostic recovery, consider the following example where we have the  initial state,
\begin{equation}
    \ket{\psi}_{\mathrm{initial}}=e^{{z\over 2N}(J_{+}^{2}-J_{-}^{2})}\ket{N,0}
    \label{eq:initial_state}
\end{equation}
where $z$ is defined according to Eq.~\eqref{eq:sq} with $\delta=0.2$ so that the state is spin-squeezed. 
Now a unitary rotation error $e^{ib J_y}$ acts on the state and we want to see how the noise agnostic recovery performs in this simple case. 
If the recovery process works, we will be able to obtain the ideal state after the recovery process.
The fidelity of the final state compared to the initial state for both in the presence and absence of the noise agnostic recovery is given in Fig.~(\ref{fig:noise_agnostic_recovery_1}).
The figure shows the fidelity of the recovered state with (\ref{eq:initial_state}) as a function of the spin $j$ ($j=N/2$) for the recovery corresponding to the syndrome with the maximum probability for $b=\sqrt{2\pi/N}/4$. 
From the figure, one can see the significant improvement in fidelity when we do the recovery operation compared to the lack of recovery, also as we go to a larger value of spin the recovered fidelity increases, whereas in the recovery-less case, the fidelity decreases and saturates to the value $e^{-(0.2)^{2}/4}$ predicted by the quantum central limit theorem.
This example indicates that the inter-ensemble interaction defining the  noise agnostic recoveries in (\ref{fig:recovery_circuit}) allows high-fidelity recovery. Similarly, to implement the noise-agnostic recovery of a shift in $J_{y}$ expectation in the second figure of (\ref{fig:recovery_circuit}), one utilizes the rotated POVM constructed from projections onto $e^{-i\pi J_z}\ket{\psi_{x}}$.
\begin{equation}
\lbrace \ket{\phi_{z}(a)}:=e^{-i\pi J_z}e^{-iaJ_{y}}e^{{z\over 2N}(J_{+}^{2}-J_{-}^{2})}\ket{N,0}\rbrace_{a=-\pi}^{\pi}.
\label{eqn:povm_p}
\end{equation}

 More generally, the full recovery channel acting on the noisy state $\mathcal{N}_{\gamma}(\rho)$ for the first part of the circuit in (\ref{fig:recovery_circuit}) is given by,
\begin{align}
    &{}\mathcal{R}^{(q)}(\mathcal{N}_{\gamma}(\rho))= \nonumber \\
    &{}   \int_{-\pi}^{\pi} da \; e^{i R_{\sqrt{2\pi/N}}(a)J_{y}} \text{tr}_{2}\left[ e^{-2i{J_x\otimes J_{y}\over N}}\mathcal{N}_{\gamma}(\rho)\otimes \ket{+_{L}}\bra{+_{L}}\right. \nonumber \\
     &{} \left. e^{2i{J_x\otimes J_{y}\over N}} \left( \mathbb{I}\otimes \ket{\psi_{z}(a)}\bra{\psi_{z}(a)} \right) \right] e^{-i R_{\sqrt{2\pi/N}}(a)J_{y}}
    \label{eqn:uuui}
\end{align}
where $\mathcal{R}^{(q)}(\mathcal{N}_{\gamma}(\rho))$ becomes normalized as $N\rightarrow \infty$ because the measurement (\ref{eqn:povm}) becomes complete in that limit. 
Recovery channels correcting for $J_{x}$ rotation errors can be defined act similar to (\ref{eqn:uuui}), and concatenating the recovery channels would lead to higher performance recovery protocols.

The noise agnostic recovery introduced in this section works in the limit of small $\delta$ and large $N$ because in those limits it becomes equivalent to the CV error correction circuit proposed in the original GKP paper \cite{PhysRevA.64.012310}. 
Note that when finite energy versions of the CV GKP codes are utilized in the recent experimental implementations, it was also necessary to adapt the original recovery method \cite{sivak2022real,PhysRevLett.125.260509,de2022error}.
One can view the finite $N$ and non-infinitesimal $\delta$ cases of the spin-$N/2$ GKP code as analogous to finite-energy versions of the CV GKP code, and in future work, it may prove advantageous to port noise-agnostic error correction schemes for finite-energy CV GKP codes to the atomic ensemble setting. 


\section{Preparation of GKP states using Linear combination of unitaries}
\label{sec:gkp_prep_LCU}
Even though CV GKP states are widely considered as  optimal states for error correction of various optical decoherence models, high-fidelity preparation protocols for these states remain a major roadblock for their use in applications. 
There have been multiple approaches for preparing the CV GKP states \cite{PhysRevA.93.012315,shi2019fault}.
In this section we discuss a general approach for preparing spin-$N/2$ GKP states using the well-known  technique of linear combination of unitaries (LCU), which is widely deployed in quantum algorithms \cite{berry2015simulating,low2019hamiltonian,holmes2022quantum}. 
The goal of the LCU is to implement a linear combination of unitaries $X=\sum_j \alpha_j U_j$, with $\alpha_j > 0$, using an ancillary qubits and entangling interaction. 
The following circuit implements the LCU,
\begin{figure}[!ht]
    \centering
    \includegraphics[width=0.42\textwidth]{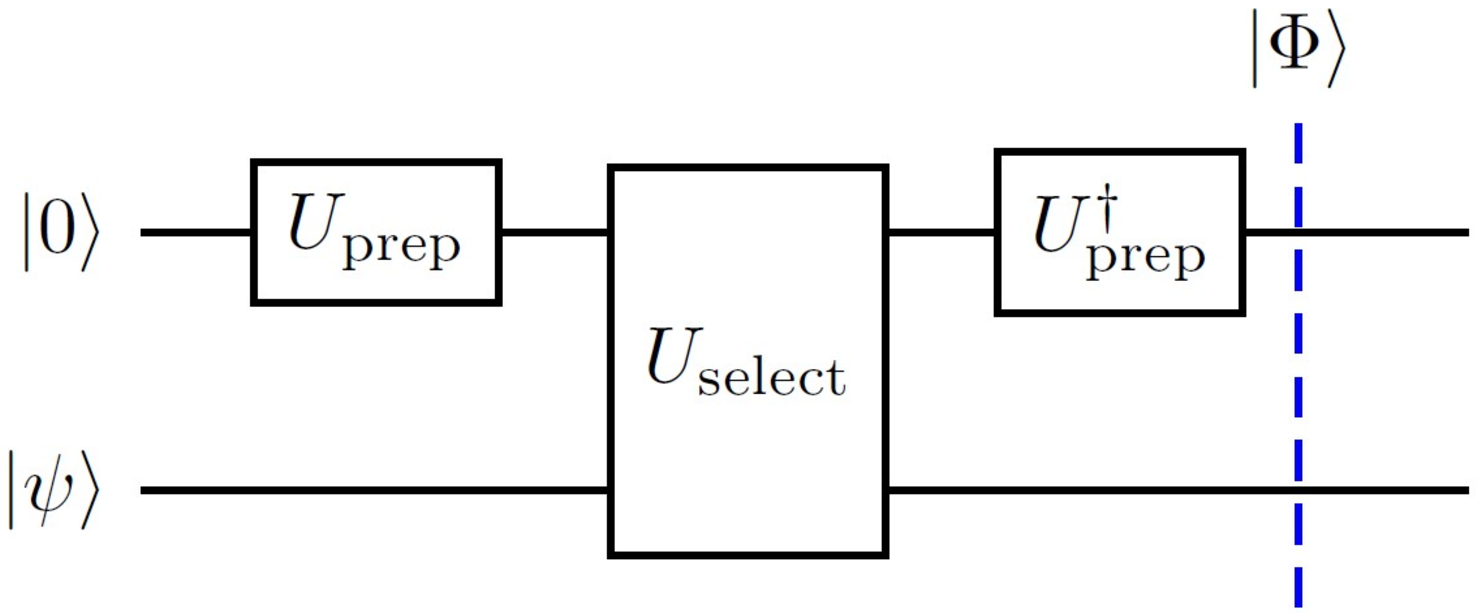}
    \label{fig:LCU_circuit}
\end{figure}
 where the action of the preparation unitary is $U_{\mathrm{prep}}\ket{0}={1\over \sum_{j}\alpha_{j}^{2}}\sum_j\alpha_j\ket{j}$ and  the select unitary defined by
 \begin{equation}
     U_{\mathrm{select}}=\sum_j\ketbra{j}{j}\otimes U_j
 \end{equation}
 implements $U_{j}$ controlled on the $\ket{j}$ state of the ancilla register, which we assume is a quantum system with discrete addressable levels. The state $\ket{\psi}$ in the above figure is the parent state for the spin-$N/2$ GKP code and is given by a pure coherent state or spin-squeezed coherent state for the code states discussed in this work.
 Thus the final state is given as,
 \begin{equation}
     \ket{\Phi}=\frac{1}{\sum_j \alpha_j^2}\ket{0}X\ket{\psi}+\ket{\phi}
 \end{equation}
 where the state $\ket{\phi}$ has only support on the space orthogonal to the ancilla state $\ket{0}$.
 Further, the oblivious amplitude amplification techniques can be used to increase the probability of heralding the desired state $X\ket{\psi}$ \cite{berry2015simulating}.
 
LCU approach is ideal for the preparation of GKP state as in general a GKP state can be written as
\begin{equation}
     \ket{\mu}=\sum_{t=-T}^{T}\alpha_{t}U(t)R_{r}(\mu)\ket{\text{VAC}}
\end{equation}
for an appropriate notion of vacuum state $\ket{\text{VAC}}$ and structure unitary $R_{r}(\mu)$ which defines the phase space squeezing structure and logical index of the code state. The coefficients $\alpha_{t}$ define the weights of the superposed nonorthogonal states according to the specific code at hand.
For example consider code states such as those for \texttt{tactgkp} and \texttt{oatgkp} which have the general form
\begin{equation}
\begin{aligned}
\ket{\mu}&=\sum_{t=-T}^{T}\alpha_{t}e^{2itJ_{y} \sqrt{2\pi\over N} } R_{r}(\mu) \ket{N,0}
\end{aligned}
\end{equation}
where the generic vacuum state has been defined as $\ket{N,0}$, the unitary $U(t)$ is now an indexed rotation around $J_{y}$, and $R_{r}(\mu)\ket{N,0}$ is the spin-squeezed state produced using TACT or OAT, respectively ($r$ is a generic parameter characterizing the squeezing).

To apply the LCU method, one can write the preparation unitary $U_{\mathrm{prep}}$ on the ancilla mode as
\begin{equation}
    U_{\mathrm{prep}}\ket{0}=\sum_{t=-T}^{T}\alpha_{t}\ket{t}
\end{equation}
and 
\begin{equation}
\begin{aligned}
 U_{\mathrm{select}}&=
 \exp\left(\frac{2i\sqrt{2\pi}}{\sqrt{N}} \mathcal{T} \otimes J_y  \right)\\
&= \sum_{t=-T}^{T}\ket{t}\bra{t}\otimes e^{2it\sqrt{2\pi \over N} J_{y}} .
\label{eq:u_select}
\end{aligned}
\end{equation}
The coupling to the ancilla in (\ref{eq:u_select}) is given here by an operator with integer spectrum $\mathcal{T}=\sum_{t=-T}^{T} t\ketbra{t}{t}$, but could be replaced by an operator with positive integer spectrum such as the photon number operator if the ancilla is a CV mode. This would require a rotation of the system at the end of the LCU circuit if a certain orientation of the code state is desired.
Application of the LCU circuit, therefore, yields
\begin{equation}
    \ket{\Phi}=\frac{1}{\sum_{t=-T}^{T}{\alpha_{t}}^2}\ket{0}\ket{\mu}+\ket{\phi}.
\end{equation}

The $U_{\mathrm{select}}$ in (\ref{eq:u_select}) has been implemented for the light-matter interaction for spin systems \cite{PhysRevLett.116.053601,PRXQuantum.3.020308} and for Bose-Einstein condensates \cite{PhysRevLett.112.233602}. 
The main limitation of the above approach is the number of accessible levels in the ancilla. 
To obtain a large number of combs in the phase space distribution for large atom number $N$ requires increasing the $T$ value, and the decoherence rate scales with the number of accessible levels in the ancilla \cite{PhysRevLett.116.053601}.
Similar techniques can be implemented for the CV GKP case  using the cross-Kerr interaction as the $U_{\mathrm{select}}$ interaction \cite{PhysRevA.105.022436}. In this case, the correct GKP state is heralded on a vacuum (no-photon) measurement result on the ancilla. 

As an illustrative example consider the case of \texttt{tactgkp(uni) } given in Eq.\eqref{eq:tact_gkp}, for this case one can find that the $U_{\mathrm{prep}}$ is proportional to a complex Hadamard matrix, viz.,
\begin{equation}
U_{\mathrm{prep}}=\frac{1}{\sqrt{2T+1}}\sum_{i,j=-T}^{T}\omega^{i j}\ketbra{i}{j}
    \label{eq:u_prep_tact_unit}
\end{equation}
where $\omega=\exp(2i \pi/(2T+1))$.

\begin{figure}
    \centering
    \includegraphics[width=0.48\textwidth]{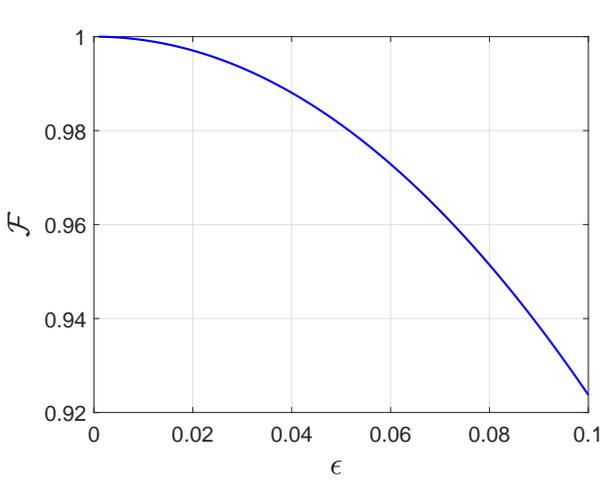}
    \caption{\textbf{Performance of the LCU approach for GKP state preparation.} 
    The fidelity of a noisily-prepared \texttt{tactgkp(uni)} to an ideal \texttt{tactgkp(uni)}  as a function of the error parameter for $T=5,\delta=0.1$ and $N=64$.  
    The fidelity remains large  even for a significant error.
    }
\label{fig:lcu}
\end{figure}

 Now as a possible source of error, one can consider coherent errors acting on the ancilla system register prior to application of $U_{\mathrm{select}}$.
 This model of errors is chosen because the entangling interaction is the hardest step of the LCU method to implement in practice. 
 Specifically, the select unitary is  replaced by the noisy select unitary
  \begin{equation}
     U_{\mathrm{select}} \to U_{\mathrm{select}} \exp(-i \epsilon H_{\mathrm{rand}}) \otimes \mathds{1}
 \end{equation}
 where $H_{\mathrm{rand}}$ is a random Hermitian operator acting on the ancilla  and we restrict the operator norm of the error-generating random Hermitian matrices to be $T$. 
Fig. \ref{fig:lcu} shows the state preparation fidelity as a function of the error parameter $\epsilon$ for the case of  $T=5$, $\delta=0.1$, and $N=64$.
 Numerically it is found that the state fidelity for the preparation of $\ket{0_{L}}$ for \texttt{tactgkp(uni)} is large even for a significant error $\epsilon$ and thus our approach is robust to imperfections.
 A similar analysis can be done for other GKP states as well showing that the state preparation is robust to imperfections.

A modification of the previous analysis allows one to use LCU to generate the codes \texttt{spingkp} and \texttt{unigkp}, which have the generic form
\begin{equation}
    \ket{\nu}=\sum_{t_1,t_2}\alpha_{t_1,t_2}U(t_1)V(t_2)\ket{N,0},
\end{equation}
for rotations $U$ and $V$.
Unlike the previous case, two ancillae are sufficient and the LCU circuit is given by
\begin{figure}[!ht]
    \centering
    \includegraphics[width=0.42\textwidth]{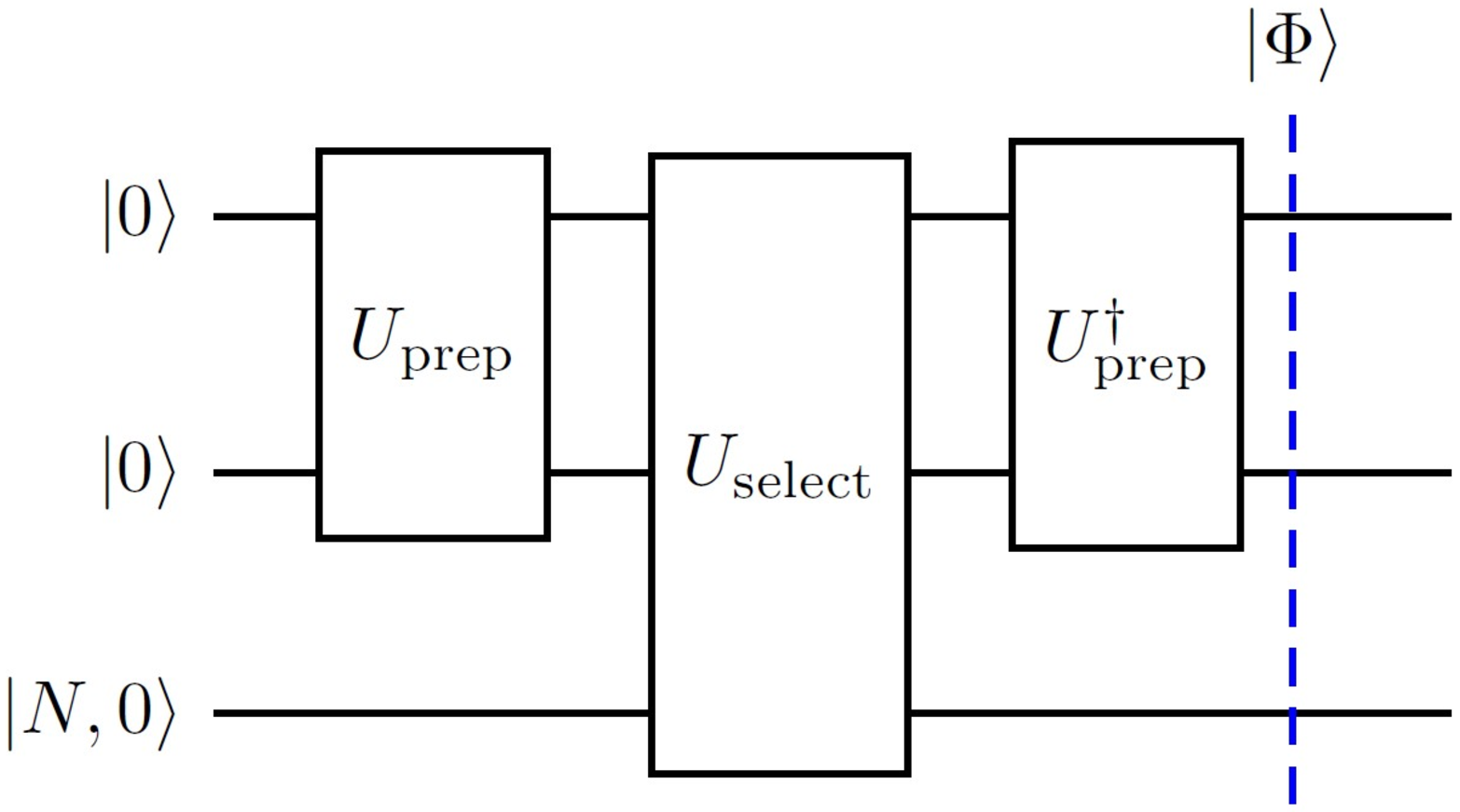}
    \label{fig:LCU_circuit_grid}
\end{figure}

where we have 
\begin{equation}
U_{\mathrm{prep}}\ket{0}\ket{0}={1\over \sum_{t_{1},t_{2}}\alpha_{t_{1},t_{2}}^{2}}\sum_{t_1}\alpha_{t_1,t_2}\ket{t_1}\ket{t_2}.
\end{equation}
The select unitary now has two control registers
\begin{equation}
    \begin{aligned}
    U_{\mathrm{select}}&=\sum_{t_{1},t_{2}}\ket{t_{1}}\bra{t_{1}}\otimes \ket{t_{2}}\bra{t_{2}}\otimes U_{t_{1}}V_{t_{2}}
    \end{aligned}
\end{equation}
 and the final state is given as,
 \begin{equation}
     \ket{\Phi}=\frac{1}{\sum_{t_{1},t_{2}} \alpha_{t_{1},t_{2}}^2}\ket{0,0}X\ket{N,0}+\ket{\phi}
 \end{equation}
 where $X=\sum_{t_{1},t_{2}}\alpha_{t_{1},t_{2}}U_{t_{1}}V_{t_{2}}$.
 
\section{Universal fault-tolerant quantum computation}
\label{sec:universal_gate_set}

The fact that the Clifford gates for the single-mode CV GKP code are given by symplectic transformations of the single-mode CV phase space $\mathbb{R}^{2}$ allows one to utilize (\ref{eq:isometric_equations}) to define approximate Clifford gates for the spin-$N/2$ GKP codes. 
Specifically, because every symplectic transformation corresponds to a CV unitary generated by a self-adjoint operator quadratic in $a$ and $a^{\dagger}$, we can write the analogue of the fault-tolerant CV GKP Clifford group generators for a spin-$N/2$ GKP code such as \texttt{tactgkp}:
\begin{equation}
    \begin{aligned}
     \mathrm{SUM}&= \exp\left(-2i{J_x\otimes J_{y}\over N}\right) \\
          F& = \exp\left(-i\frac{\pi}{2}J_z\right),\\
          P&= \exp \left( i{J_{x}^{2}\over N}\right).
    \end{aligned}
\end{equation}
The approximate stabilizer generators are
\begin{align}
        \exp\left(-2i\sqrt{\frac{2\pi}{N}}J_x\right) \text{ and } \exp\left(2i\sqrt{\frac{2\pi}{N}}J_y\right)
\end{align} where the choice of rotation generators can be inferred from the phase space distributions of the code states shown in Fig. \ref{fig:husimi_code_words}. The $\mathrm{SUM}$ gate for two spin-$N/2$ systems requires access to an all-to-all coupling between the individual atoms of the two subsystems. 
For example, in ion traps the availability of all-to-all couplings provides a path to implement the SUM gate using the M{\o}lmer-Sorenson interaction  \cite{PhysRevLett.82.1971,PhysRevLett.82.1835}.
Also, in a Rydberg atom system in which coherent
transport of entangled atom arrays has recently been demonstrated, this technique could be used to implement the $\mathrm{SUM}$ gate \cite{bluvstein2022quantum}.
It may also be possible to extend methods for generating spin squeezing in a single atomic ensemble \cite{PhysRevA.66.022314,PhysRevLett.116.053601,PRXQuantum.3.020308}, to a scheme in which total spin operators of two atomic ensembles are coupled by off-resonant interaction with a single laser. Such an approach would not require single-atom addressability, but may not be able to enter a strongly-interacting regime.

A non-Clifford resource is required to make a universal gate set for quantum computing.
One approach is to utilize a magic state and, analogous to the method proposed in \cite{PhysRevLett.123.200502}, one can create an H-type magic state for the spin-$N/2$ GKP code by initializing a spin-$N/2$ coherent state (analog of the CV vacuum state) and using the GKP error correction protocol using the $\mathrm{SUM}$ gate a state of the distillable H-type magic state \cite{PhysRevA.71.022316,reichardt2005quantum}.

The main difference between the fault-tolerant gates for the continuous variable case to the spin case we defined here is that all the gates for the continuous variable case are generated by symplectic interactions whereas, for the case of the spin case, we need $\mathrm{SU(2)}$ plus one non-linear interaction for the $P$ gate.

\subsection{Propagation of errors}
To show the feasibility of the fault-tolerant gates for computation with spin-$N/2$ GKP states, one needs to determine whether the gates amplify errors present in a specific implementation\cite{PhysRevX.10.011058}. 
For simplicity, we focus on coherent rotation errors,
\begin{equation}
    \begin{aligned}
        M_{\theta}&=\exp(-i\theta J_x),\\
        N_{\phi}&=\exp(-i\phi J_y).
    \end{aligned}
\end{equation}

Now consider the $F$ gate, we get that,
\begin{equation}
    \begin{aligned}
        F M_{\theta} F^{\dagger}&= \exp\left(-i\frac{\pi}{2}J_z\right) M_{\theta} \exp\left(i\frac{\pi}{2}J_z\right)\nonumber \\
        &= N_{-\theta} ,
    \end{aligned}
\end{equation}
and similarly
\begin{equation}
    F N_{\phi} F^{\dagger}=\exp(-i\phi J_x)=M_{\phi}.
\end{equation}
Concatenating the noise-agnostic recovery channels $\mathcal{R}^{(q)}$, $\mathcal{R}^{(p)}$ of Section \ref{sec:nar} allows to correct for small $J_x$ and $J_y$ errors, so the error propagation is controllable.

Next consider the gate$P$, it commutes with $M_{\theta}$ and the effect of $P$ on $N_{\phi}$ is  given as,
\begin{equation}
    \begin{aligned}
        P N_{\phi} P^{\dagger}&= \exp \left( i{J_{x}^{2}\over N}\right) N_{\phi} \exp \left( -i{J_{x}^{2}\over N}\right).
    \end{aligned}
\end{equation}
Using the canonical commutation relationship between the continuous variables $q$ and $p$, and the isometry $V$ from (\ref{eq:isometric_equations}) we get for large $N$,
\begin{align}
     Ve^{i{J_{x}^{2}\over N}} e^{-i \phi J_{y}} e^{-i{J_{x}^{2}\over N}}V^{\dagger}&\sim e^{iq^{2}/2}e^{-i\phi \sqrt{N\over 2}p} e^{-iq^{2}/2}\nonumber \\
     V N_{\phi}M_{-\phi}V^{\dagger}&\sim e^{-i\phi \sqrt{N\over 2}p}e^{i\phi\sqrt{N\over 2}q}.
\end{align}
But the operators on the right-hand sides of these equations are equal up to an overall phase. We conclude that for large $N$, the $N_{\phi}$ error propagates through the $P$ gate by introducing $N_{\phi}M_{-\phi}$ errors, which are 
again correctable by concatenation of the  recovery channels $\mathcal{R}^{(q)}$, $\mathcal{R}^{(p)}$.

Next, first noting that  certain local $M_{\theta}$ and $N_{\phi}$ errors on the SUM gate are benign due to  
\begin{equation}
    \begin{aligned}
        \mathrm{SUM}\left( M_{\theta} \otimes \mathds{1} \right)\mathrm{SUM}^{\dagger}&= 0,\\
         \mathrm{SUM}\left( \mathds{1}  \otimes N_{\phi}\right)\mathrm{SUM}^{\dagger}&= 0,\\
    \end{aligned}
\end{equation}
we find that 
\begin{align}
   V\mathrm{SUM}\left( N_{\phi} \otimes \mathds{1} \right)\mathrm{SUM}^{\dagger}V^{\dagger} \sim e^{-i\sqrt{N\over 2}(p\otimes \mathbb{I}+\mathbb{I}\otimes p)}\nonumber \\
   V N_{\phi}^{\otimes 2}V^{\dagger}\sim e^{-i\phi \sqrt{N\over 2}p}\otimes e^{-i\phi \sqrt{N\over 2}p} 
\end{align}
so the effect of the $N_{\phi}$ error on the SUM gate is to fan out the error to another atom ensemble register. If the other register is also a spin-$N/2$ GKP qubit, the local errors are individually correctable.  An analogous conclusion holds for the $M_{\phi}$ error.

The final ingredient in the fault-tolerant quantum computation scheme is magic state preparation, which involves initial preparation of a spin coherent state followed by an error correction scheme composed of SUM gates. Therefore, the possibility of controlling error propagation in the SUM gate indicates the possibility of fault-tolerant magic state preparation, at least for coherent rotation errors.
Thus all the gadgets in the fault-tolerant scheme for the spin-$N/2$ GKP code propagate these errors in a controllable manner.

\section{Discussion}
\label{sec:discussions}
 CV GKP codes are considered as primary candidates for implementing fault-tolerant, scalable quantum computation because of the natural robustness of these codes to errors in CV bosonic systems. 
 In this paper, we utilized spin squeezing interactions and the quantum central limit theorem to develop a notion of GKP codes for atomic ensembles, which are candidates for large scale quantum information processing with ultracold matter. 
In particular, our code states exhibit a well-defined comb-like phase space structure similar to the CV GKP states, and the free parameters defining the comb-like structure can be optimized to obtain high optimal recovery rates under the stochastic relaxation noise and ballistic dephasing.
The stochastic relaxation channel is analogous to the CV amplitude damping process and the isotropic ballistic dephasing noise is analogous to CV Gaussian thermal noise.
For the stochastic relaxation channel, the \texttt{spinGKP} code is the spin-$N/2$ GKP code allowing the best optical recovery, and it further outperforms the standard bosonic codes even for intermediate noise strengths. For isotropic ballistic dephasing, \texttt{oatgkp}, which presently has no well-studied CV GKP analogue, outperforms other bosonic codes for intermediate noise strength values. 
We studied how to use techniques from quantum algorithms to create the spin-$N/2$ GKP states, as there is no fundamental atomic interaction that generates the GKP state.
We used the linear combination of unitaries (LCU) to prepare the spin-$N/2$ GKP code states with the only controlled being rotations. This approach is in principle adaptable to CV GKP and other non-trivial state preparations which are superpositions of Gaussian states or approximately Gaussian states with fixed interaction strength. 
 Finally, we again used the quantum central limit theorem to outline the correspondence between a gate set allowing universal, fault-tolerant computation with CV GKP-encoded qubits and a gate set with generators given by total spin operators of a spin-$N/2$ system. For sufficiently large $N$, these gates provide a route to universal, fault-tolerant quantum computation using a spin-$N/2$ GKP-encoded qubits.



In this work, we have focused on atomic ensembles which can be prepared in $\mathrm{SU}(2)$ coherent states, which are relevant when the atoms are distributed among two orthogonal modes (e.g., internal or motional). Similar codes can be defined for $\mathrm{SU}(d)$ coherent states, which are relevant in the case of cold atoms with a larger number of accessible internal states or distributed in an optical trapping lattice\cite{smith2013,omanakuttan2022qudit,omanakuttan2021quantum,zache2023fermion}. 
Analyses of such GKP codes for extended ultracold matter are an important direction for future studies. We also expect that concatenation of spin-$N/2$ GKP codes with standard surface code schemes potentially provides a route to fault-tolerant quantum computation with ultracold atoms,  analogous to proposals for incorporating CV GKP codes in larger QEC schemes. Detailed threshold analyses and comparison to state-of-the-art atomic gates are necessary to substantiate this claim as a realistic proposal.

Lastly, we note that spin-$N/2$ GKP codes are not the only quantum error correction codes for atomic ensembles in the symmetric subspace. Such symmetric logical qubits have previously been considered as  possible candidates to achieve fault-tolerant quantum computation \cite{PhysRevA.90.062317,PhysRevA.93.042340,omanakuttan2023multispin}.
In \cite{PhysRevA.90.062317}, the ground state space of Heisenberg ferromagnets was used as physical motivation for introducing permutation-invariant codes for ensembles of $N$ atoms. 
In contrast, the present work was motivated by the  relationship between quadrature squeezing in infinite-dimensional CV systems and spin squeezing in the symmetric subspace of $N$ two-level atoms.
This relationship is made rigorous by the QCLT, which we further utilized in translating noise-agnostic error correction protocols and fault-tolerant gates from the CV GKP setting. Because the relevant spin-squeezing interactions have already been implemented experimentally, and because several methods for utilizing CV GKP codes for fault-tolerant measurement-based CV quantum computation have now been proposed \cite{Bourassa2021,PhysRevA.101.012316,PRXQuantum.3.010315,Larsen2021,PhysRevX.8.021054,PhysRevA.107.052414}, it is interesting to consider the possibility of implementing spin-$N/2$ GKP codes in schemes for fault-tolerant quantum computation in atomic ensembles.

\acknowledgements

The authors thank Victor Albert for discussions about bosonic codes and optimal recovery methods as well as other useful discussions. S.O. would like to acknowledge fruitful discussions with Tyler G Thurtell, Austin K Daniel, and Karthik  Chinni during various stages of work.  T.J.V. acknowledges support from the Laboratory Directed Research and Development (LDRD) Program at Los Alamos National Laboratory (LANL).
S.O. acknowledges the support by the Laboratory Directed
Research and Development program of Los Alamos National Laboratory under Project No. 20200015ER and by the U.S. Department of Energy, Office of Science, National Quantum Information Science Research Centers, Quantum Systems Accelerator.
LANL is operated by Triad National Security, LLC, for the National Nuclear Security Administration of U.S. Department of Energy (Contract No. 89233218CNA000001).

\appendix
\section{Diversity combining method for optimal recovery}
\label{sec:app_diversity_combining}

Here we provide an overview of our implementation of the diversity combining method for finding the recovery channel that maximizes the channel fidelity \cite{PhysRevA.75.012338}. Let the logical code states be elements of a $d$-dimensional Hilbert space $\mathcal{H}$. The diversity combining method offers a quadratic advantage in the size of the semidefinite program (from $d^{4}$ to $4d^{2}$ optimized parameters) compared to a standard optimization of the entanglement fidelity because the logical states are replaced by qubit states. Consider the ensemble of code states: $\rho={1\over 2}\ket{0_{L}}\bra{0_{L}}+{1\over 2}\ket{1_{L}}\bra{1_{L}}$, where we assume $\langle 0_{L}\vert 1_{L}\rangle =0$ (if the code states are not orthogonal, use the Gram-Schmidt procedure to define orthonormal code states $\ket{0_{L}}$ and $\ket{1_{L}}$). The encoding isometry is defined as $S=\ket{0_{L}}\bra{0}+\ket{1_{L}}\bra{1}$.  We consider the following error channel:
\begin{equation}
\mathcal{N}_{\gamma}\circ \mathcal{S}\circ \mathcal{S}^{\dagger}(\rho) = \mathcal{N}_{\gamma}(SS^{\dagger}\rho SS^{\dagger}).
\label{eqn:mapmap}
\end{equation}
Since $SS^{\dagger}=P_{\text{code}}$, the above error map is equal to $\mathcal{N}_{\gamma}(\rho)$ (note that $P_{\text{code}}\rho P_{\text{code}}=\rho$). Note that
\begin{equation}
\rho ' := \mathcal{S}^{\dagger}(\rho)=S^{\dagger}\rho S={1\over 2}\ket{0}\bra{0}+{1\over 2}\ket{1}\bra{1}
\end{equation}
is a 2$\times$2 matrix. The map $\mathcal{N}_{\gamma}\circ \mathcal{S}$ has Kraus operators $\lbrace E_{\ell}S\rbrace_{\ell}$, where each Kraus operator is a $d\times 2$ matrix, and one can see that the map in (\ref{eqn:mapmap}) is $\mathcal{N}_{\gamma}\circ \mathcal{S}(\rho ')$.  We implement recovery for $\mathcal{N}_{\gamma}\circ \mathcal{S}$ by a recovery channel $\mathcal{R}$ with Kraus operators $R_{j}$ which are $2\times d$ matrices (so $\mathcal{R}$ takes the noisy encoded state back to the qubit space). Then any purification $\ket{\psi_{\rho '}}$ of $\rho '$ satisfies
\begin{equation}
\langle \psi_{\rho'}\vert (\mathcal{R}\circ \mathcal{N}_{\gamma}\circ \mathcal{S}) \otimes \mathbb{I} \left( \ket{\psi_{\rho'}}\bra{\psi_{\rho'}} \right) \vert \psi_{\rho'} \rangle = \sum_{j,\ell}\vert \text{tr} R_{j}E_{\ell}S \rho' \vert^{2}.
\label{eqn:rbrb1}
\end{equation}

The latter expression is expressed as an inner product of Choi matrices by using the $\textbf{vec}$ map from $\mathcal{H}\otimes \mathcal{H}^{\text{dual}}$ to $\mathcal{H}^{\otimes 2}$ given by $\textbf{vec}E_{i,j}= \vert E_{i,j}\rangle\rangle :=\ket{i}\ket{j}$, where $E_{i,j}:= \ket{i}\bra{j}$. Write (\ref{eqn:rbrb1}) as

\begin{align}
\sum_{j,\ell}\vert \text{tr} R_{j}E_{\ell}S \rho' \vert^{2}&= \sum_{j,\ell} \langle\langle \rho' \vert R_{j}E_{\ell}S \rangle\rangle \langle \langle R_{j}E_{\ell}S\vert  \rho ' \rangle\rangle \nonumber \\
 &{}= \sum_{j,\ell} \langle \langle \rho ' \vert (\mathbb{I}\otimes S^{T}E_{\ell}^{T})\vert R_{j}\rangle\rangle \langle\langle R_{j}\vert (\mathbb{I}\otimes E_{\ell}^{*}S^{*}) \vert \rho '  \rangle\rangle \nonumber  \\
&= \sum_{\ell} \langle \langle \rho ' S^{\dagger}E_{\ell}^{\dagger}\vert J_{\mathcal{R}} \vert \rho ' S^{\dagger}E_{\ell}^{\dagger} \rangle\rangle\nonumber \\
&= \text{tr}C_{\rho ',\mathcal{N}_{\gamma}}J_{\mathcal{R}}
\end{align}

with $C_{\rho ',\mathcal{N}_{\gamma}}:= \sum_{\ell}\vert \rho ' S^{\dagger}E_{\ell}^{\dagger} \rangle\rangle\langle \langle \rho ' S^{\dagger}E_{\ell}^{\dagger}\vert$. 
Note that $J_{\mathcal{R}}$ is a $2d\times 2d$ Hermitian matrix and that $\text{tr}C_{\rho ',\mathcal{N}_{\gamma}} = 1/2$.

Therefore, the maximal channel fidelity
\begin{align}
F_{\mathcal{N}_{\gamma}}=\max_{\mathcal{R}}\langle \psi_{\rho'}\vert (\mathcal{R}\circ \mathcal{N}_{\gamma}\circ \mathcal{S}) \otimes \mathbb{I} \left( \ket{\psi_{\rho'}}\bra{\psi_{\rho'}} \right) \vert \psi_{\rho'} \rangle \nonumber
\end{align}
is given by the solution to the semidefinite program
\begin{equation}
\begin{aligned}
& \underset{X}{\text{maximize}}
& & \text{tr}(C_{\rho ',N_{\gamma}}X) \\
& \text{subject to}
& & X  >0 \\
&&& \text{tr}_{1}X=\mathbb{I}_{d}
\end{aligned}
\label{eqn:rtyrt}
\end{equation}
\begin{equation}
\begin{aligned}
& \underset{Y}{\text{minimize}}
& & \text{tr}(Y) \\
& \text{subject to}
& & -C_{\rho',\mathcal{N}_{\gamma}} + \mathbb{I}_{2}\otimes Y  >0 
\end{aligned}
\end{equation}
over $d\times d$, Hermitian $Y$.

Note that if one considers the Kraus operators $R_{s,j}:= \lbrace {1\over \sqrt{2}}\ket{s}\bra{j}\rbrace_{s=1,2;j=1,\ldots, d}$ for the recovery channel $\mathcal{R}$, one has $\sum_{s,j}R_{s,j}^{\dagger}R_{s,j}=\sum_{s,j}\text{tr}_{1}\vert R_{s,j}\rangle\rangle \langle\langle R_{s,j}\vert = \mathbb{I}_{d}$, as required. Since $\sum_{s,j}\text{tr}_{1}\vert R_{s,j}\rangle\rangle \langle\langle R_{s,j}\vert  ={1\over 2}\mathbb{I}_{2} \otimes \mathbb{I}_{d}$, it follows that the maximum value of the above semidefinite program is at least $1/4$. Physically, one can see that this lower bound is attained by considering the identity noise channel and the recovery that randomly sends each of the orthogonal code states to $\ket{0}$, $\ket{1}$ with equal probability.

\section{Properties of stochastic relaxation channel}
\label{app:sbp}
We have the stochastic relaxation channel given as in \eqref{eqn:oppp}:
\begin{equation}
    T_{t}(\rho)=\sum_{\ell=0}^{N}{(1-e^{-\lambda t})^{\ell}\over \ell!\lambda^{\ell}}e^{-{\lambda t\over 4} (N-2J_{z})}(\mathcal{S}^{(1)})^{\ell}(\rho)e^{-{\lambda t\over 4} (N-2J_{z})},
    \label{eq:stochastic_relaxation}
\end{equation}
where we have defined,
\begin{align}
    &{}\mathcal{S}^{(1)}(\ket{N-m,m}\bra{N-m',m'})\nonumber \\
    &{} =\lambda J_{+}{\ket{N-m,m}\bra{N-m',m'}  \over \sqrt{(N-m+1)(N-m'+1)}}J_{-}  .
\end{align} 
To show that this map is trace-preserving, it suffices to verify that $\text{tr}\left[ T_{t}(\ket{N-m,m}\bra{N-m',m'})\right] = \delta_{m,m'}$. For $m\neq m'$, it is clear that $\text{tr}\left[ T_{t}(\ket{N-m,m}\bra{N-m',m'})\right]=0$ because 
\begin{align}
&{} J_{+}^{\ell}\ket{N-m,m}\bra{N-m',m'}J_{-}^{\ell} \propto \nonumber  \\&{} \ket{N-m+\ell,m-\ell}\bra{N-m'+\ell,m'-\ell}.
\end{align}
For $m=m'$, one has 
\begin{widetext}
\begin{align}
    \text{tr}\left[ T_{t}(\ket{N-m,m}\bra{N-m,m})\right]&= \text{tr}\left[ \sum_{\ell=0}^{N}{(1-e^{-\lambda t})^{\ell}\over \ell!(N-m+1)\cdots(N-m+\ell)}e^{-{\lambda t\over 4} (N-2J_{z})}J_{+}^{\ell}\ket{N-m,m}\bra{N-m,m} J_{-}^{\ell}e^{-{\lambda t\over 4} (N-2J_{z})} \right] \nonumber \\
    &=\text{tr}\left[ \sum_{\ell=0}^{N}{(1-e^{-\lambda t})^{\ell}\over \ell!}{m!\over m-\ell!}e^{-{\lambda t\over 4} (N-2J_{z})}\ket{N-m+\ell,m-\ell}\bra{N-m+\ell,m-\ell} e^{-{\lambda t\over 4} (N-2J_{z})} \right]\nonumber \\
    &= \text{tr}\left[ \sum_{\ell=0}^{N}(1-e^{-\lambda t})^{\ell}{m \choose \ell}e^{-\lambda t(m-\ell)}\ket{N-m+\ell,m-\ell}\bra{N-m+\ell,m-\ell} \right]\nonumber \\
    &=  \sum_{\ell=0}^{m}(1-e^{-\lambda t})^{\ell}{m \choose \ell}e^{-\lambda t(m-\ell)}\nonumber \\
    &= e^{-\lambda t m} \sum_{\ell=0}^{m}{m \choose \ell}(e^{\lambda t}-1)^{\ell}\nonumber \\
    &=1.
\end{align}

We now consider the $\ell=0$ and $\ell=1$ contributions to the channel in the limit of small $\gamma:=t\lambda$ and find that up to linear order in $\gamma$, one has $\sum_{\ell=0}^{1}E_{\ell}^{\dagger}E_{\ell} = \mathbb{I}$, where 
\begin{equation}
    \begin{aligned}
E_{0}&=  \begin{pmatrix}
  1 &    \\
  & 1-{\gamma\over 2} &   \\
  & & 1-{2 \gamma\over 2}&  \\
  & & & \ddots & \\
  & & & & 1- {N\gamma\over 2}
 \end{pmatrix}\nonumber,
E_{1}&= \sqrt{\gamma}
 \begin{pmatrix}
  0 & 1 &   \\
  & 0 & \sqrt{2} &   \\
  & & \ddots & \ddots & \\
  & & & 0 & \sqrt{N}  \\
  & & & & 0
 \end{pmatrix}.
\end{aligned}
\end{equation}
\end{widetext}
are matrices on the $N+1$ dimensional symmetric subspace. Embedding the symmetric subspace into two CV oscillator modes according to $\ket{N-m,m}\mapsto \ket{N-m}\otimes \ket{m}$, one can write
  \begin{align}
E_{0}&= \mathbb{I}-\frac{\gamma}{2}\mathbb{I}_{1}\otimes a_{2}^{\dagger}a_{2} \nonumber\\
E_{1}&= \sqrt{\gamma }R\otimes a_{2}.
\end{align} 
where $a_{1}, a_{2}$ are the respective annihilation operators of the CV modes and $R$ is the unilateral shift $R\ket{n}=\ket{n+1}$.
Therefore, for small $\gamma$, one can think of the channel $T_{t}$ as an attenuation of the second mode which still conserves the global total particle number.

\section{Numerical Benchmarking details}
\label{sec:Numerical_Benchmarking_details}
\begin{table}
\centering

 \begin{tabular}{ |c|c|c|c|c| } 
 \hline
 $\gamma$ & \texttt{spingkp} ($\delta$)& \texttt{oatgkp} ($\delta$)  & \texttt{tactgkp}($\delta$)&\texttt{unigkp} ($T$)\\
\hline 
  0.0100    &0.2926  &  0.0777&    0.2715&   3\\
    0.0303  &  0.3055&    0.0785 &   0.2746 &  3\\
    0.0505   & 0.3110&    0.0790 &   0.2746 &  3\\
    0.0708    &0.3096&    0.0794 &   0.2732 &  3\\
    0.0910    &0.3067&    0.0797 &   0.2717 &  3\\
    0.1113 &   0.3035  &  0.0799  &  0.2698  &  3\\
    0.1315  &  0.3002 &   0.0801 &   0.2678 &   3\\
    0.1518  &  0.2970 &   0.0803 &   0.2657 &   2\\
    0.1721&    0.2938 &   0.0804 &   0.2657&   2\\
    0.1923 &   0.2909 &   0.0805 &   0.2628 &   2\\
    0.2126 &   0.2882 &   0.0806 &   0.2618 &   2\\
    0.2328 &   0.2858 &   0.0807 &   0.2613 &   2\\
    0.2531 &   0.2838 &   0.0807&    0.2616 &   2\\
    0.2733  &  0.2824 &   0.0807 &   0.2627 &   2\\
    0.2936 &   0.2818 &   0.0807 &   0.2650 &   2\\
    0.3138 &   0.2821 &   0.0807&    0.2687 &   2\\
    0.3341 &   0.2835 &   0.0807&    0.2739 &   2\\
    0.3544 &   0.2862 &   0.0807&    0.2815 &   2\\
    0.3746 &   0.2904 &   0.0808&    0.2922 &   2\\
    0.3949 &   0.2963 &   0.0808&    0.3084 &   2\\
    0.4151 &   0.3045 &   0.0808&    0.3349 &   2\\
    0.4354 &   0.3153 &   0.0808&    0.3725 &   2\\
    0.4556 &   0.3301 &   0.0808&    0.3991 &   1\\
    0.4759 &   0.3509 &   0.0808&    0.3999 &   1\\
    0.4962 &   0.3816 &   0.0809&    0.3999 &   1\\
    0.5164 &   0.4000 &   0.0809&    0.3999 &   1\\
    0.5367 &   0.4000 &   0.0809 &   0.4000 &   1\\
    0.5569 &   0.4000 &   0.0810 &   0.4000 &   1\\
    0.5772 &   0.4000 &   0.0811 &   0.4000 &   1\\
    0.5974 &   0.4000 &   0.0814 &   0.4000 &   1\\
    0.6177 &   0.4000 &   0.0817 &   0.4000  &  1\\
    0.6379 &   0.4000 &   0.0821 &   0.4000 &   1\\
    0.6582 &   0.4000&    0.0825 &   0.4000 &   1\\
    0.6785 &   0.4000&    0.0829 &   0.4000  &  1\\
    0.6987 &   0.4000 &   0.0834&    0.4000 &   1\\
    0.7190 &   0.4000&   0.0839&    0.4000&    1\\
    0.7392 &   0.4000 &   0.0843 &   0.4000 &  1\\
    0.7595 &   0.4000&    0.0848 &   0.4000 &  1\\
    0.7797 &   0.4000&    0.0852 &   0.4000&   1\\
    0.8000 &   0.4000&    0.0855 &   0.4000&    1\\
\hline
\end{tabular}
    \caption{The optimized code parameters for the stochastic relaxation channel using diversity combining approaches for the data Fig.~\ref{fig:stochastic_relaxation} for $N=64$.
    For the case of the \texttt{unigkp} we optimize over various choices of the $1\leq T\leq 5$ and the best possible $T$ is chosen.
    For the codes $\{\texttt{spingkp}, \texttt{oatgkp}, \texttt{tactgkp}\}$ we choose $T=5$, and   for the case of the \texttt{spingkp} and \texttt{tactgkp} we restrict the value of $\delta$ to be with $0\leq \delta \leq 1$ whereas for the case of the $\texttt{oatgkp}$ the value of $\delta$ is restricted to $0\leq \delta \leq 1/\sqrt{N}$.
     The restrictions are made such that the code states retain the comb-like structure essential for the GKP decoding process and the optimal value occurs approximately when the two code words become orthogonal as shown in Fig.~(\ref{fig:fidelity_figure}). }
    \label{tab:stochastic_relaxation}
\end{table}
The value of the parameters that optimize the channel fidelity $F_{\mathcal{E}}$ for the stochastic relaxation channel data in Fig.~\ref{fig:stochastic_relaxation} is given in Table \ref{tab:stochastic_relaxation}.
The codewords we are interested here are are the $\{\texttt{spingkp}, \texttt{oatgkp}, \texttt{tactgkp}, \texttt{unigkp}\}$.
For the case of the \texttt{unigkp} we optimize over various choices of the $1\leq T\leq 5$ and the best possible $T$ is chosen.
    For the codes $\{\texttt{spingkp}, \texttt{oatgkp}, \texttt{tactgkp} \}$ we choose $T=5$, and   for the case of the \texttt{spingkp} and \texttt{tactgkp} we restrict the value of $\delta$ to be with $0\leq \delta \leq 1$ whereas for the case of the $\texttt{oatgkp}$ the value of $\delta$ is restricted to $0\leq \delta \leq 1/\sqrt{N}$.
     The parameter domain restrictions are made such that the comb-like structure of the phase space support of the code states persists, which is essential for the GKP decoding process.

\begin{table}
\centering

 \begin{tabular}{ |c|c|c|c|c|} 
 \hline
 $\sigma$ & \texttt{spingkp}($\delta$) & \texttt{oatgkp}($\delta$)  & \texttt{tactgkp}($\delta$) & \texttt{unigkp}($T$)\\
\hline 
0.0010 & 0.2200 & 0.0682&0.2200  &3\\
    0.0012& 0.2200&0.0682&0.2200&3\\
    0.0014&0.2200&0.0682&0.2200&3\\
    0.0017&0.2200&0.0682&0.2200&3\\
    0.0020&0.2200&0.0682&0.2200&3\\
    0.0024&0.2200&0.0682&0.2200&3\\
    0.0029&0.2200&0.0682&0.2200&3\\
    0.0035&0.2200&0.0682&0.2200&3\\
    0.0041&0.2200&0.0682&0.2200&3\\
    0.0049&0.2200&0.0682&0.2200&3\\
    0.0059&0.2200&0.0682&0.2200&3\\
    0.0070&0.2200&0.0682&0.2200&3\\
    0.0084&0.2200&0.0682&0.2200&3\\
    0.0100&0.2196&0.0682& 0.2200 &3\\
    0.0119&0.2183&0.0682&0.2196&3\\
    0.0143&0.1926&0.0681&0.1926&3\\
    0.0170&0.2200&0.0682&0.2049&3\\
    0.0203&0.2201&0.0668&0.2178&3\\
    0.0242&0.1925&0.0629&0.2023&3\\
    0.0289&0.2116&0.0682&0.2062&3\\
    0.0346&0.2201&0.0597&0.2113&3\\
    0.0412&0.2187&0.0673&0.2171&3\\
    0.0492&0.2199&0.0665&0.2199&3\\
    0.0588&0.2178&0.0511&0.2178&3\\
    0.0702&0.2199&0.0544&0.2199&3\\
    0.0838&0.2193&0.0683&0.2308&3\\
    0.1000&0.2199&0.0685&0.2299&3\\
    0.1194&0.2525&0.0683&0.2194&3\\
    0.1425&0.2157&0.0683&0.2199&3\\
    0.1701&0.2393&0.0683& 0.2289&2\\
    0.2031&0.2428&0.0630&0.2202&3\\
    0.2424&0.2135&0.0709&0.2173&2\\
    0.2894&0.2362&0.0702&0.2141&1\\
    0.3455&0.2221&0.0513&0.1980&1\\
    0.4125&0.1799&0.0707&0.1652&1\\
    0.4924&0.1909&0.0751&0.1752&1\\
    0.5878&0.1925&0.0731&0.1652&1\\
    0.7017&0.1925&0.0751&0.1864&1\\
    0.8377&0.1652&0.0676&0.1787&1\\
    1.000&0.2205& 0.0682&0.2013&1\\
\hline
\end{tabular}
    \caption{ The optimized parameters for the isotropic ballistic dephasing channel using diversity combining approaches given in Fig.~\ref{fig:ballistic_isotropic} for $N=64$.
    For the case of the \texttt{spingkp} we restrict the value of $\delta$ to be with $0\leq \delta \leq 1$ whereas for the case of the $\texttt{oatgkp}$ the value of $\delta$ is restricted to $0\leq \delta \leq 1/\sqrt{N}$. 
    For both cases, we studied the case of $T=5$.
     The restrictions are made such that the code states retain the comb-like structure essential for the GKP decoding process and the optimal value occurs approximately when the two code words become orthogonal as shown in Fig.~(\ref{fig:fidelity_figure}). }
    \label{tab:isotropic_ballistic}
\end{table}
The value of the parameters that optimizes the $F_{\mathcal{E}}$ for the isotropic ballistic dephasing channel in Fig.~\ref{fig:ballistic_isotropic} is given in Table \ref{tab:isotropic_ballistic}.
The codewords we are interested here are are the $\{\texttt{spingkp}, \texttt{oatgkp},\texttt{tactgkp}\}$.
For the case of the \texttt{spingkp} we restrict the value of $\delta$ to be with $0\leq \delta \leq 1$ whereas for the case of the $\texttt{oatgkp}$ the value of $\delta$ is restricted to $0\leq \delta \leq 1/\sqrt{N}$.
  For both cases, we studied the case of $T=5$.
     The restrictions are made such that we are in a regime where we still have the comb-like structure essential for the GKP decoding process

\section{Ballistic dephasing along one axis\label{sec:bd}}
\label{sec:Ballistic_dephasing_along_one_axis}
A one-axis ballistic dephasing, which describes a Hamiltonian with fluctuating energy levels and studied in detail in \cite{baamara2021squeezing} is defined by the following mixed-unitary channel in Eq.~(\ref{eqn:ballistic_dephasing}).
For $\sigma \ll 1$, the effect of this channel can be seen by considering, e.g., the observable $J_{x}$, and noting that
\begin{align}
    \text{Var}_{\Phi_{\sigma}(\rho)}J_{x} = (1-\sigma)\text{Var}_{\rho}J_{x} + \sigma \langle J_{y}^{2}\rangle_{\rho} + O(\sigma^{2}).
\end{align}

Consider the case when the dephasing occurs about the $z$-axis and write the Kraus operators as,
\begin{equation}
    E_{\theta}=e^{-\frac{\theta^2}{4\sigma}} e^{-i\theta J_{z}}.
\end{equation}
When $\sigma$ is taken to be small, one can focus on the $\theta\approx 0$ limit. In this limit, the quantum error correction condition
depends on only the first and second moments of the total spin operator $J_z$.
To identify codes that have large value of the correctable part of the QEC kernel (\ref{eqn:qeckern}) in the small $\sigma$ limit, we expand about $\theta, \theta'=0$:
\begin{equation}
\begin{aligned}
P_{\mathrm{code}}E_{\theta}^{\dagger}E_{\theta'}P_{\mathrm{code}}&= e^{-\frac{\theta^2+\theta'^2}{4\sigma}} P_{\mathrm{code}}\\
&-\frac{1}{2}e^{-\frac{\theta^2+\theta'^2}{4\sigma}}(\theta-\theta')^2 P_{\mathrm{code}}J_z^2 P_{\mathrm{code}}\\
&+\mathcal{O}(\theta^3)+\mathcal{O}(\theta'^3),
\label{eq:qec_ballistic}
\end{aligned}
\end{equation}
where $ P_{\mathrm{code}}=\ketbra{0_{L}}{0_{L}}+\ketbra{1_{L}}{1_{L}}$ and we have assumed that the code under consideration has no first $J_z$ moment, i.e., $P_{\mathrm{code}}J_zP_{\mathrm{code}}=0$.
Codes for which the code states are not connected by pair tunneling between single-particle modes will retain a diagonal matrix in the second line of Eq.\eqref{eq:qec_ballistic}, indicating greater correctability.
Examples include the coherent state codes $\ket{J,\vec{n}\cdot \vec{J}=\frac{N}{2}}$
where $\vec{n}$ is a unit vector in the $xy$-plane, or cat codes along direction $\vec{n}$.

\begin{figure*}
    \centering
    \includegraphics[width=\textwidth]{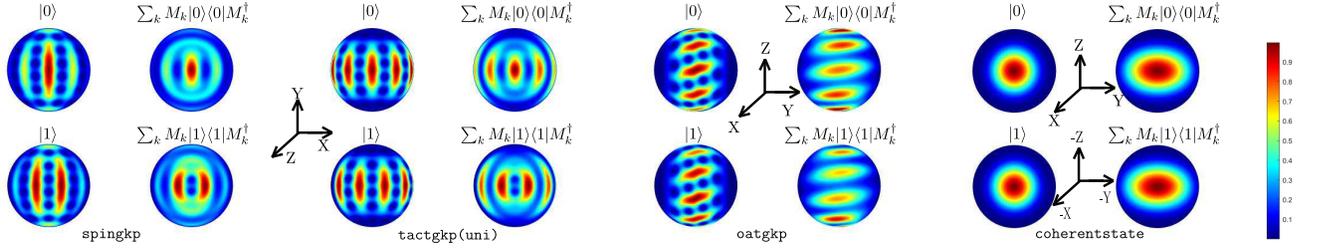}
    \caption{\textbf{Husimi distribution for ballistic dephasing.} Husimi distribution of the codewords for the \texttt{oatgkp} and \texttt{spingkp} for the quantum channel considered in Eq.~(\ref{eqn:ballistic_dephasing}) for $\gamma=0.2$. 
    The plots are shown for $\delta=0.2$ and $\delta=0.04$ respectively for the \texttt{spingkp}  and \texttt{oatgkp}.
    The effect of the channel for these two codewords are similar in nature for the \texttt{spingkp} and \texttt{oatgkp}. We can see the effect of this quantum channel for these codewords is to make the codewords more delocalized in phase space; while still maintaining the comb-like structure. The persistence of these comb-like structures makes the recovery accurate for both these codes. But the spreading of the comb structure makes the fidelity decay faster compared to the stochastic relaxation case.  }
    \label{fig:ballistic_dephasing_husimi}
\end{figure*}

\begin{figure}
    \centering
    \includegraphics[width=\columnwidth]{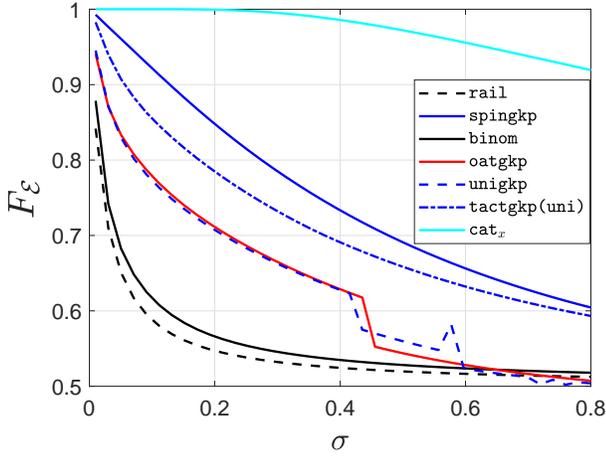}
    \caption{\textbf{Channel fidelity for ballistic dephasing.} The channel fidelity as a function of $\sigma$ for the noise model $\mathcal{E}=\Phi_{\sigma}$ given in Eq.~\eqref{eqn:ballistic_dephasing}. 
    The ballistic dephasing along $z$ direction is given for different code words with $N=64$ and cut off, $T=5$ for all codes except \texttt{unigkp} codes where we all optimize for all values of $T$ . 
    The figure shows that the GKP codes outperform the binomial code and rail code for large range of $\sigma$, however the cat state is the best as predicted from the QEC matrix calculation in Eq.\eqref{eq:qec_ballistic}. 
     }
    \label{fig:ballistic_dephasing_jx_jz}
\end{figure}

We now examine the optimal recovery properties of spin-$N/2$ GKP codes under  one-axis ballistic dephasing.
In Fig.~\ref{fig:ballistic_dephasing_husimi}, the effect of this channel for the case of dephasing along $z$ direction for $\gamma=0.2$ is shown. The effect of the noise channel is similar for the \texttt{spingkp} and \texttt{oatgkp}, in particular, the code states have more delocalized support along an axis in phase space. However, the comb-like structure is maintained, leading to high optimal recovery for both these codes. The spreading of the comb structure causes the fidelity to decay faster compared to the stochastic relaxation case.
Fig.~\ref{fig:ballistic_dephasing_jx_jz} shows the optimal recovery performance of various spin-$N/2$ codes with respect to the one-axis ballistic dephasing channel for $N=64$. A cutoff $T=5$ is used for all codes except \texttt{unigkp} codes where we all optimize for all values of $T$. The figure shows that the spin-$N/2$ codes outperform \texttt{binomial} and \texttt{rail} for a large range of $\sigma$, however, \texttt{cat} has the highest optimal recovery performance as predicted from calculation in \eqref{eq:qec_ballistic}. 

\section{\label{sec:app} Optimal recovery for CV GKP codes}
\begin{figure*}
\includegraphics[width=0.9\textwidth]{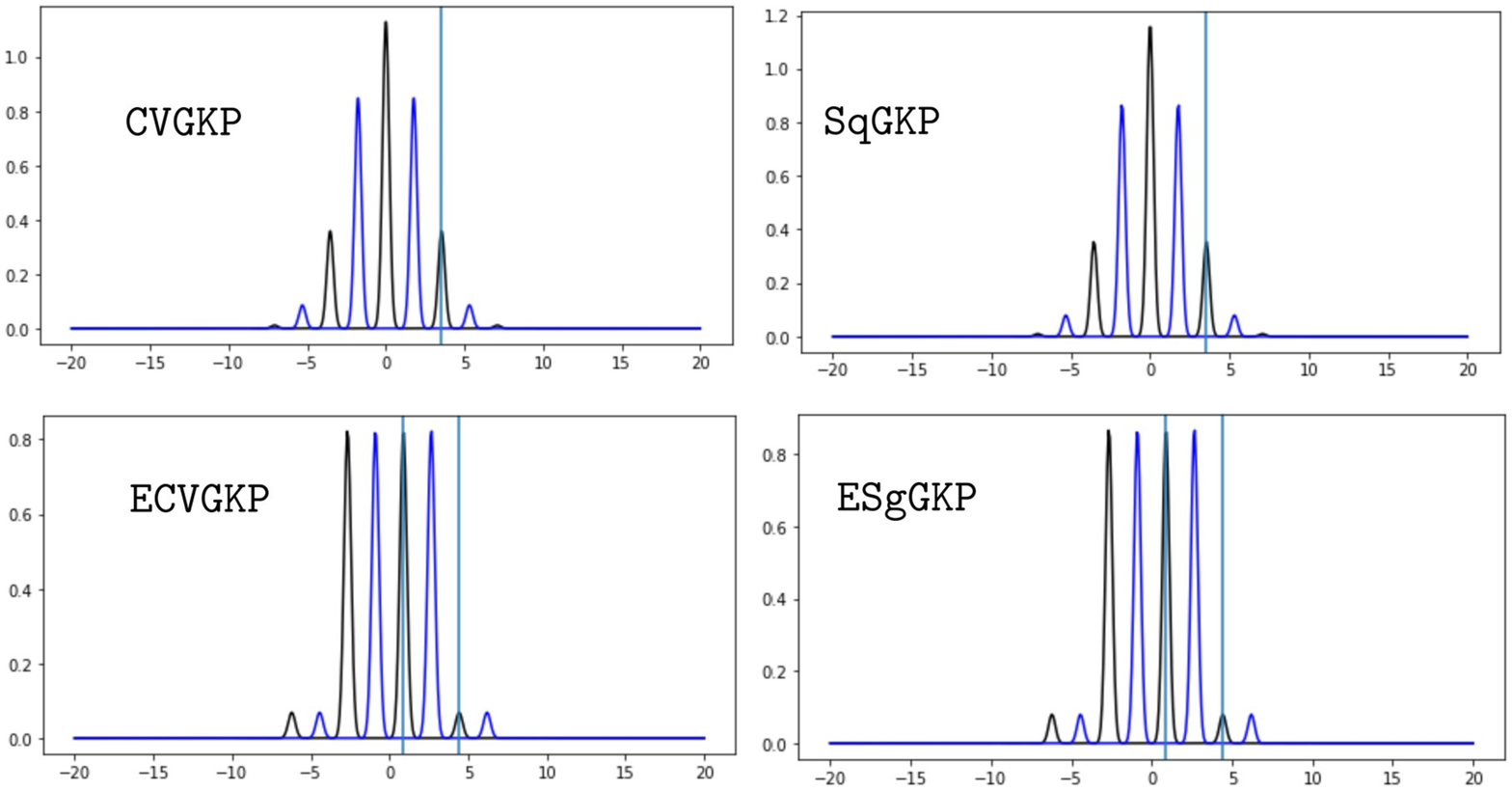}
    \caption{
    The position space wavefunctions for the code states for all  four codes are given.
From the figure, one can see the well-separated comb-like structure for all these four codewords.
For the equal energy codewords, the  position space peaks for the logical $\ket{0}_\mathrm{L}$ and $\ket{1}_\mathrm{L}$ are in one-to-one correspondence.
The square grid equal energy code is taken to have energy $5$ and $T=8$; the vertical lines are shown at ${\sqrt{\pi}\over 2}$ and ${\sqrt{\pi}\over 2}+2\sqrt{\pi}$ for reference.
}
    \label{fig:gfc5}
\end{figure*}
There is not a unique finite-energy regularization scheme for the CV GKP code, and a specific convention can be chosen with regard to accessible experimental state preparation schemes \cite{PhysRevA.102.032408}.
However, at a fixed energy the finite-energy GKP codes can have different optimal recovery properties. 
We define four different kinds of finite energy CV GKP codes. First, we consider Gottesman's original finite energy code \texttt{CVGKP},
\begin{equation}
\begin{aligned}
\vert \mu_{\lambda , d}\rangle &\propto \sum_{t\in \mathbb{Z}}e^{-{\pi\over 2} \lambda^{2}(2t+\mu)^{2}}\\
&{}D\left(\sqrt{\pi\over 2}(2t+\mu)\right)S\left({1\over 2}\log d\right)\vert 0\rangle
\end{aligned}
\end{equation}
$\mu \in \lbrace 0,1\rbrace$.
Often one takes $d={1\over \lambda^{2}}$, so that $\lambda$ is the only parameter.
However, one can also choose $d$ based on fidelity minimization or best optimal recovery.  In practice, the infinite sum over $t$ is truncated to the domain $[-T,T]$ for integer $T$.

Another finite energy CV GKP code is defined by a superposition of CV coherent states on a square lattice in $\mathbb{R}^{2}$. We refer to this code as \texttt{SgGKP},
\begin{equation}
\begin{aligned}
\vert \mu_{\lambda}\rangle \propto \sum_{t\in \mathbb{Z}^{\times 2}}&e^{-{\pi\over 2} \lambda^{2}((2t_{1}+\mu)^{2} + t_{2}^{2})}\\
&D\left(\sqrt{\pi\over 2}(2t_{1}+\mu)\right)D\left(i\sqrt{\pi\over 2}t_{2}\right)\vert 0\rangle,
\end{aligned}
\label{eqn:sqgridcv}
\end{equation}
where $\mu \in \lbrace 0,1\rbrace$.

We can also define CV GKP codes for which the code states have exactly equal energy in expectation. Analogous to \texttt{CVGKP}, one defines the code \texttt{ECVGKP} as,
\begin{equation}
\begin{aligned}
\vert \nu_{\lambda , d}\rangle \propto \sum_{t\in \mathbb{Z}}&e^{-2\pi \lambda^{2}t^{2}}\\
&D\left(\sqrt{\pi\over 2}(2t+\nu)\right)S\left({1\over 2}\log d\right)\vert 0\rangle
\end{aligned}
\end{equation}
$\nu \in \lbrace -{1\over 2},{1\over 2}\rbrace$. 
There is also an equal expected energy version of \texttt{SgGKP} which we call \texttt{ESgGKP}. It is given by
\begin{equation}
\begin{aligned}
\vert \nu_{\lambda}\rangle \propto \sum_{t\in \mathbb{Z}^{\times 2}}&e^{-{\pi\over 2} \lambda^{2}((2t_{1}+2\nu)^{2} + t_{2}^{2})}\\
&D\left(\sqrt{\pi\over 2}(2t_{1}+\nu)\right)D\left(i\sqrt{\pi\over 2}t_{2}\right)\vert 0\rangle
\end{aligned}
\end{equation}
$\nu \in \lbrace -{1\over 2},{1\over 2}\rbrace$.

In Fig.~\ref{fig:gfc5}, the position space wavefunctions for each of these four codes are shown. From the figure, one can identify a well-defined comb-like structure for all codewords.
Also for the codes \texttt{SgGKP} and call \texttt{ESgGKP} the position space peaks have the same amplitudes for the logical $\ket{{1\over 2}_{L}}$ and $\ket{-{1\over 2}_{L}}$.

\begin{figure*}
    \centering
    \includegraphics[width=\textwidth]{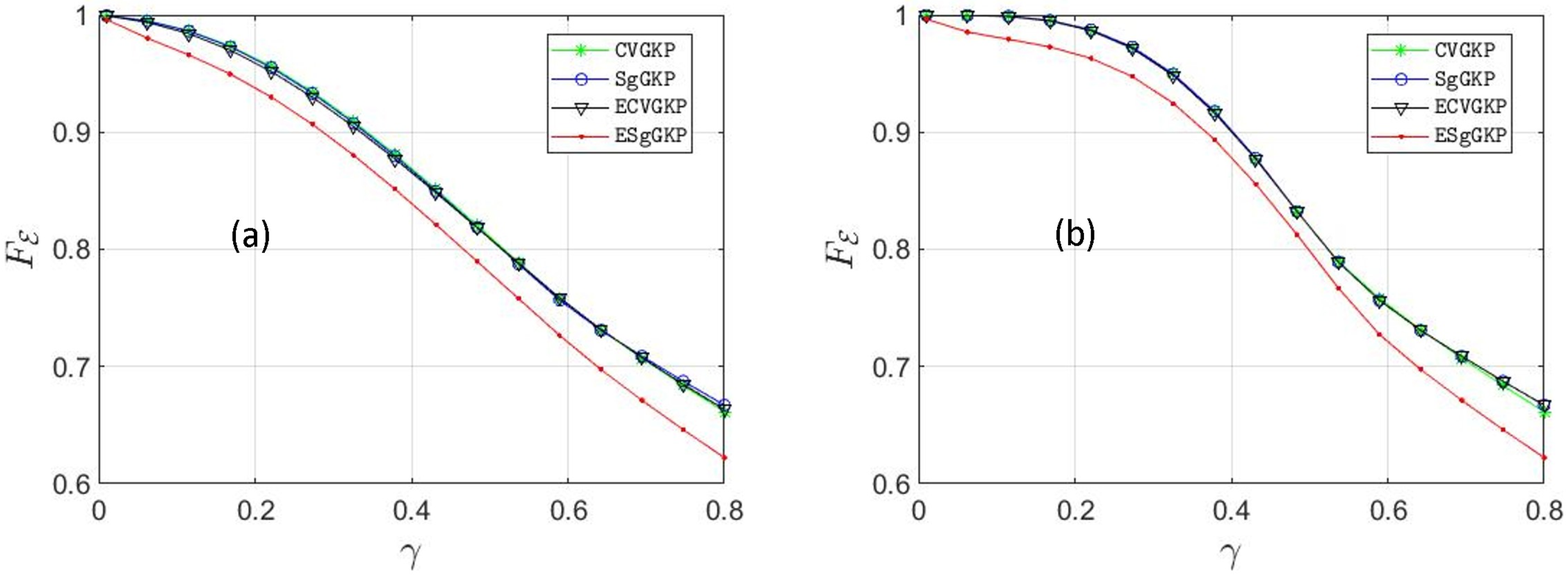}
    \caption{The optimal recovery performance of finite-energy CV GKP codes under the quantum channel given in Eq.~(\ref{eq:photon_detection_process}). 
    In (a) the case of $T=2$ is shown whereas in (b) the case of $T=5$ is shown.}
    \label{fig:compare_cv_gkp}
\end{figure*}
Similar to the case of the spin-$N/2$ GKP states, one can use the diversity combining method and optimization of the code parameters  to compare the optimal recovery  of the four finite-energy CV GKP codes above   (Fig.~\ref{fig:compare_cv_gkp}). The noise channel is given by \eqref{eq:photon_detection_process}. In (a) the case of $T=2$ is shown whereas in (b) the case of $T=5$ is shown.
From these curves, it is clear that the equal expected energy version of the square grid CV GKP code has lower optimal achievable fidelity. This suggests that given a specific noise channel, it is important to consider an appropriate finite-energy CV GKP code to maximize recovery.



\bibliography{gkpbib.bib}
\end{document}